\documentclass[fleqn,usenatbib]{mnras}

\usepackage{newtxtext,newtxmath}

\usepackage[T1]{fontenc}
\usepackage{ae,aecompl}

\usepackage{graphicx}	
\usepackage{amsmath}	
\usepackage{amssymb}	
\usepackage{booktabs}
\usepackage{makecell}
\usepackage{multirow}

\usepackage{etoolbox}
\makeatletter
\patchcmd\@combinedblfloats{\box\@outputbox}{\unvbox\@outputbox}{}{%
  \errmessage{\noexpand\@combinedblfloats could not be patched}%
}%
\makeatother

\newcommand{\UV}{\mbox{$\mathrm{UV_{275}}$}}
\newcommand{\U}{\mbox{$\mathrm{U_{336}}$}}
\newcommand{\B}{\mbox{$\mathrm{B_{438}}$}}
\newcommand{\V}{\mbox{$\mathrm{V_{606}}$}}
\newcommand{\Vwfpc}{\mbox{$\mathrm{V_{555}}$}}
\newcommand{\I}{\mbox{$\mathrm{I_{814}}$}}
\newcommand{\kev}{\mathrm{keV}}
\newcommand{\cory}[1]{{\color{black}{#1}}}
\newcommand{\coryriv}[1]{{\color{black}{#1}}}
\newcommand{\toreferee}[1]{{\color{black}{#1}}}
\newcommand{\msun}{\mbox{$M_\odot$}}
\newcommand\nodata{ ~$\cdots$~ }

\title[Faint X-ray sources in M30]{A deep {\it Chandra} survey for faint X-ray sources in the Galactic globular cluster M30, and searches for optical and radio counterparts}

\author[Y. Zhao et al.]{
Yue Zhao,$^{1}$\thanks{E-mail: zhao13@ualberta.ca}
Craig O. Heinke,$^{1}$
Haldan N. Cohn,$^{2}$
Phyllis M. Lugger,$^{2}$
\newauthor
Sebastien Guillot,$^{3,4}$
Constanza Echibur\'u,$^{5}$ 
Laura Shishkovsky,$^{6}$
Jay Strader,$^{6}$
\newauthor
Laura Chomiuk,$^{6}$
Arash Bahramian,$^{7}$
James C. A. Miller-Jones,$^{7}$
\newauthor
Thomas J. Maccarone,$^{8}$
Evangelia Tremou,$^{9}$
and Gregory R. Sivakoff$^{1}$
\\
$^{1}$Department of Physics, University of Alberta, CCIS 4-183, Edmonton, AB T6G 2E1, Canada\\
$^{2}$Department of Astronomy, Indiana University, 727 E. Third St. Bloomington, IN 47405, USA\\
$^{3}$IRAP, CNRS, 9 avenue du Colonel Roche, BP 44346, F-31028 Toulouse Cedex 4, France\\
$^{4}$Universit\'e de Toulouse, CNES, UPS-OMP, F-31028 Toulouse, France\\
$^{5}$Department of Physics, McGill University, 3600 rue University, Montreal, QC, Canada, H3A 2T8\\
$^{6}$Center for Data Intensive and Time Domain Astronomy, Department of Physics and Astronomy, Michigan State University, \\
East Lansing, MI 48824, USA\\
$^{7}$International Centre for Radio Astronomy Research-Curtin University, GPO Box U1987, Perth, WA 6845, Australia\\
$^{8}$Department of Physics \& Astronomy, Texas Tech University, Box 41051, Lubbock, TX 79409-1051, USA\\
$^{9}$LESIA, Observatoire de Paris, CNRS, PSL, SU/UPD, Meudon, France
}

\date{Accepted XXX. Received YYY; in original form ZZZ}

\pubyear{2019}

\begin{document}
\label{firstpage}
\pagerange{\pageref{firstpage}--\pageref{lastpage}}
\maketitle

\begin{abstract}
We present a deep ($\sim 330~\mathrm{ks}$) {\it Chandra} survey of the Galactic globular cluster M30 (NGC 7099). Combining the new Cycle 18 with the previous Cycle 3 observations we report a total of 10 new X-ray point sources within the $1\farcm 03$ half-light radius, compiling an extended X-ray catalogue of a total of 23 sources. We incorporate imaging observations by the {\it Hubble Space Telescope} and the {\it Karl G. Jansky Very Large Array} from the MAVERIC survey to search for optical and radio counterparts to the new and old sources. Two X-ray sources are found to have a radio counterpart, including the known millisecond pulsar PSR J2140$-$2310A, the radio position of which also matches a previously reported faint optical counterpart which is slightly redder than the main sequence. We found optical counterparts to $18$ of the $23$ X-ray sources, identifying $2$ new cataclysmic variables (CVs), $5$ new CV candidates, $2$ new candidates of RS CVn type of active binary (AB), and $2$ new candidates of BY Dra type of AB. The remaining unclassified X-ray sources are likely background active galactic nuclei (AGN), as their number is consistent with the expected number of AGN at our X-ray sensitivity. Finally, our analysis of radial profiles of different source classes suggests that bright CVs are more centrally distributed than faint CVs in M30, consistent with other core-collapsed globular clusters.
\end{abstract}




\section{Introduction}
Globular clusters (GCs) are old and dense stellar populations and have been intensively studied because of their interesting dynamical features. Specifically, X-ray missions with enhanced angular resolutions and instrumental sensitivity (e.g., {\it Chandra X-ray Observatory}) have revealed that GCs harbour an overabundance of point-like X-ray sources. Attributed to the very dense core of GCs, these sources are generally thought of as close binaries. Many of these close binaries  
originate from the close few-body encounters 
in the dense cores of clusters 
\citep[e.g.,][]{Fabian75, Hills76, camilo2005, ivanova2006, Ivanova08}.

The most well-known class of X-ray emitting binaries are low-mass X-ray binaries (LMXBs), which involve neutron stars (NSs) or black holes (BHs) in close orbits with low-mass companions. The NS systems are further classified into systems with relatively persistent X-ray luminosities, vs.  transient systems; the latter stay mostly in the quiescent state (\toreferee{with typical $L_X\sim 10^{31-33}~\mathrm{erg~s^{-1}}$}), known as quiescent low-mass X-ray binaries (qLMXBs) and occasionally exhibit outbursts with luminosities typically $\sim 10^4$ times brighter (see \citealt{Galloway17} for a recent review). 
LMXBs are considered to be the progenitors of millisecond radio pulsars (MSPs), 
where the NS has been spun up (``recycled") to rotate at millisecond periods by accreting matter from the companion, eventually turning on as a radio pulsar \citep{Bhattacharya91}. Quiescent LMXBs involving BHs, on the other hand, are generally fainter in X-rays (\toreferee{$L_X\sim 10^{30-33}~\mathrm{erg~s^{-1}}$}), but emit strong radio emission via their synchrotron-emitting jets
\citep{Fender03,Plotkin13,Gallo18}, enabling identification by their radio-to-X-ray flux ratio \citep{Maccarone05,Strader12,MillerJones15}.

Other X-ray emitting close binaries in GCs include cataclysmic variables (CVs) and chromospherically active binaries (ABs), which dominate  the faint ($L_X \lesssim 10^{33}~\mathrm{erg~s^{-1}}$) X-ray populations. CVs are white dwarfs (WDs) accreting from low-mass companions, typically seen at $10^{30} <L_X<10^{33}~\mathrm{erg~s^{-1}}$ in GCs \citep{hertz1983, cool1995, pooley2002}; these systems are usually identified through discoveries of optical/UV counterparts or strong optical variability (see e.g., \citealt{cohn10,RiveraSandoval18}). ABs are tidally locked close binaries that involve either main sequence (BY Dra) or evolved (RS CVn) stellar components. Their X-rays are thought to originate from active coronal regions induced by fast rotations as a result of tidal synchronisation; a dozen or more are present in many globular clusters at $L_X\sim10^{30}~\mathrm{erg~s^{-1}}$ \citep{bailyn1990, dempsey1993, grindlay2001, heinke2005b, cohn10}. 

Deep radio continuum observations of GCs have recently become possible with bandwidth upgrades to the Australia Telescope Compact Array ({\it ATCA}) and the Karl G. Jansky Very Large Array ({\it VLA}). These have enabled the detection of numerous radio point sources in GCs. Many are background active galactic nuclei (AGN), as expected from deep blank-field radio number counts \citep{Kellermann08,Padovani09}. MSPs have steep radio spectra \citep{Kramer99}, and are seen in large numbers in GCs\footnote{For a summary of MSPs in GCs, see \url{http://www.naic.edu/~pfreire/GCpsr.html}}, usually via pulsed emission \citep{Lyne87,Camilo00,camilo2005,Wang2020}, but also in continuum imaging \citep{Hamilton85,Fruchter00,McConnell01,Zhao20}. Several faint, flat-spectrum radio sources have recently been identified in GCs as black hole candidates \citep{Strader12,Chomiuk13,MillerJones15,Bahramian17,Tudor18}, though some may be other objects such as unusual NS LMXBs \citep[][a candidate transitional millisecond pulsar]{Bahramian18}, or exotic binaries consisting of normal stars and/or white dwarfs \citep{Shiskovsky18}. 

M30 (NGC 7099) is a core-collapsed GC \citep{Djorgovski86, Lugger95} at a distance of $8.1~\mathrm{kpc}$ (\citealt{harris1996}; 2010 edition). 
A previous study of a $50~\mathrm{ks}$ {\it Chandra} observation by \citet{Lugger07}  presented a catalogue of $13$ ($\lesssim 10^{33}~\mathrm{erg~s^{-1}}$) X-ray sources within M30's half-mass radius ($1\farcm15$), plus optical counterparts identified from observations with the {\it Hubble Space Telescope} ({\it HST}). \citet{Echiburu20} reported analyses of new, deep, Cycle 18 {\it Chandra} observations of the bright qLMXB in M30, \coryriv{which is dedicated to constraining the mass and radius of the NS}. 
In this work, we incorporate results from X-ray ({\it Chandra}), radio ({\it VLA}) and optical ({\it HST}) observations on M30, cataloguing and identifying faint X-ray sources by searching for possible optical and/or radio counterparts. The paper is organised as follows: In Section \ref{sec:observations}, we describe the observational data used in this work and relevant reduction procedures; in Section \ref{sec:data_analyses}, we present methodologies of our data analyses; in Section \ref{sec:discussion_individual_src}, we provide discussions on individual sources based on our results; and finally, in Section \ref{sec:conclusions}, we summarise the results and draw conclusions.


\section{Observations and data reduction}
\label{sec:observations}
\subsection{{\it Chandra} observations}
M30 has been visited in Cycle 3 (Obs. ID 2679; PI: Cohn) and Cycle 18 (PI: Guillot) by {\it Chandra} with the ACIS-S camera, totaling $\sim 330$ ks of exposure. The Cycle 18 observations were performed in the {\it very faint} (VFAINT) mode to optimise background cleaning. To reduce frame time, and therefore pileup of the bright qLMXB in the core, a 1/8 subarray centred on the cluster was used. As a result, the Cycle 18 FOV does not add more exposure to most of the cluster outskirts, which were more completely covered by the Cycle 3 observation. The analyses in this work thus 
focus on faint sources within or close to the half-light radius ($1\farcm03$; \citealt[][2010 edition]{harris1996}). More details of the observations are listed in Table \ref{tab:x_ray_obs}.

The level-1 ACIS data products are first reprocessed and aligned to the up-to-date calibration (CALDB 4.8.2) using the {\tt chandra\_repro} task in the {\sc chandra interactive analysis of observations ({\sc ciao})}\footnote{\url{http://cxc.cfa.harvard.edu/ciao/}} software \citep{Fruscione06}. This renders level-2 event files that can be used to generate scientific products for further analyses. 

\begin{table}
    \centering
    \caption{{\it Chandra} observations of M30.}
    \resizebox{\columnwidth}{!}{
    \begin{tabular}{ccccc}
    \toprule
     Cycle  & Obs. ID & Exposure (ks) & Start of Obs & Instrument \\
    \midrule
       $3$  & $2679$  &    $49.43$    & 2001-11-19 02:55:12 & ACIS-S \\
    \midrule
       $18$ & $20725$ &    $17.49$    & 2017-09-04 16:33:05 & ACIS-S \\
       $18$ & $18997$ &    $90.19$    & 2017-09-06 00:05:19 & ACIS-S \\
       $18$ & $20726$ &    $19.21$    & 2017-09-10 02:09:13 & ACIS-S \\
       $18$ & $20732$ &    $47.90$    & 2017-09-14 14:23:17 & ACIS-S \\
       $18$ & $20731$ &    $23.99$    & 2017-09-16 18:04:17 & ACIS-S \\
       $18$ & $20792$ &    $36.86$    & 2017-09-18 04:21:43 & ACIS-S \\
       $18$ & $20795$ &    $14.33$    & 2017-09-22 11:39:56 & ACIS-S \\
       $18$ & $20796$ &    $30.68$    & 2017-09-23 06:09:30 & ACIS-S \\
    \bottomrule
    \end{tabular}
    }
    \label{tab:x_ray_obs}
\end{table}

\subsection{{\it HST} observations}
We use imaging data observed by the {\it HST} Wide-field Camera 3 (WFC3; GO-13297; PI: Piotto) and the Advanced Camera for Surveys (ACS; GO-10775; PI: Sarajedini). These provide images of excellent sub-arcsecond spatial resolution in F275W ($\UV$), F336W ($\U$), and F438W ($\B$) for WFC3, and in F606W ($\V$) and F814W ($\I$) for ACS. All imaging products are composed of single exposures that are pipe-lined, flat-fielded and have charge transfer efficiency (CTE) trails removed (FLC images). The two WFC3 observations are separated roughly by 2 months (Table \ref{tab:hst_obs}), each of which contains one ($\B$) and two ($\UV$ or $\U$) dithered exposures. The latter observations are roughly rotated by $\approx 90^\circ$ relative to the earlier observations. Observations by ACS were done rougly $8$ years earlier than the WFC3 observations, of which each filter ($\V$ or $\I$) comprises one short ($7$ s) and $4$ long ($140$ s each) exposures. A summary of basic information about these {\it HST} observations is presented in Table \ref{tab:hst_obs}.

These datasets have been fully analysed as part of the {\it ACS Globular Cluster Treasury Program} (\citealt{sarajedini2007,
anderson2008})\footnote{\url{https://archive.stsci.edu/prepds/acsggct}} and the {\it Hubble Space Telescope UV Legacy Survey of Galactic Globular Clusters} (HUGS; \citealt{piotto2015}), offering a set of kinematic and 5-band photometry information. Although we generate our own photometry catalogues for optical identifications, the HUGS data products provide useful information in cases where stars are missed by our photometry (see Section \ref{sec:optical_photometry}) and are especially important in determining cluster membership \citep{Nardiello18}.

\begin{table*}
    \centering
    \caption{{\it HST} observations of M30.}
    \begin{tabular}{ccccc}
    \toprule
       GO   & Exposure (s) & Start of Obs & Instrument/Channel & Filter  \\
    \midrule
    $10775$ & $567$  & 2006-05-02 21:47:08 &  ACS/WFC  & F606W (\V) \\
    $10775$ & $567$  & 2006-05-02 23:21:19 &  ACS/WFC  & F814W (\I) \\
    \midrule
    $13297$ & $1450$ & 2014-06-08 22:13:14 & WFC3/UVIS & F275W (\UV) \\
    $13297$ & $1450$ & 2014-08-19 04:53:35 & WFC3/UVIS & F275W (\UV) \\
    $13297$ & $606$ & 2014-06-08 22:05:34 & WFC3/UVIS & F336W (\U)  \\
    $13297$ & $606$ & 2014-08-19 05:16:51 & WFC3/UVIS & F336W (\U)  \\
    $13297$ & $65$  & 2014-06-08 22:01:50 & WFC3/UVIS & F438W (\B)  \\
    $13297$ & $65$  & 2014-08-19 03:56:58 & WFC3/UVIS & F438W (\B)  \\
    \bottomrule
    \end{tabular}
    \label{tab:hst_obs}
\end{table*}

\subsection{{\it VLA} observations}
M30 was observed by {\it VLA} as a part of the ``{\it Milky Way {\it ATCA} and {\it VLA} Exploration of Radio sources In Clusters}" (MAVERIC) survey (\citealt{Tremou18}, Shishkovsky, L., et al. 2020, submitted). The observations (NRAO/VLA Program IDs: 15A-100; PI Strader) were performed in three separate blocks, on 2015-06-23, 2015-07-04, and 2015-07-05, totaling $8~\mathrm{hr}$ on source. In each block, M30 was observed with the most extended A configuration, using the C band ($4$-$8~\mathrm{GHz}$) receivers. The data were taken using the 3-bit mode, with two separate $2048~\mathrm{MHz}$ bands centered on $4.9$ and $7~\mathrm{GHz}$, respectively. Data reduction and imaging were done with {\sc aips} \citep{Greisen03} and {\sc casa} \citep{McMullin07}, rendering root mean square (RMS) values for $4.9~\mathrm{GHz}$ and $7~\mathrm{GHz}$ of $1.7~\mathrm{\mu Jy/beam}$ and $1.6~\mathrm{\mu Jy/beam}$, respectively; correspondingly the synthesized beam sizes are $0.74\arcsec \times 0.39\arcsec$ and $0.52\arcsec \times 0.28\arcsec$ at $4.9~\mathrm{GHz}$ and $7~\mathrm{GHz}$, respectively. 

We then generate radio source catalogue using the processed images. More details on the methodologies and the catalogue will be presented in a separate paper (Shishkovsky, L., et al. 2020, submitted), while in this work, we only report the radio sources that positionally match sources in our updated X-ray catalogue.

\section{Data analyses}
\label{sec:data_analyses}
\subsection{Merging the X-ray observations}
\label{sec:merge_chandra_obs}
To detect faint X-ray sources, typically to differentiate very faint sources from background fluctuations, it is important to take full advantage of all exposures. Prior to combining the files, we refine the relative WCS information for each event file to account for instrumental offsets. These offsets might be on sub-arcsecond scales, but are crucial for later detections of very faint sources. For this purpose, we chose the longest {\it Chandra} observation (Obs. ID 18997) as the reference frame, to which we calculate relative offsets for all other observations using the centroid positions of A1, the brightest source. The resulting shifts are then used as  input to the {\sc ciao} {\tt wcs\_update} and the {\tt acis\_process\_events} tools to update the WCS information for each event file. The latter generates updated event files while preserving the Energy-Dependent Subpixel Event Repositioning (EDSER; \citealt{Li04}) pixel adjustment. In a final step, we run the {\tt merge\_obs} script\footnote{
on the stack of WCS-corrected files. This tool first reprojects all input files to a common tangent point, and then creates a merged event file while generating a combined exposure map and an exposure-corrected X-ray flux image. 

We applied a $2\farcm5\times 2\farcm5$ square spatial filter and energy filters to the merged event file, creating X-ray images binned to a quarter of an arcsec (i.e., half the ACIS pixel size) over a soft ($0.5$--$2~\kev$), a hard ($2$--$7~\kev$), and a broad ($0.5$--$7~\kev$) energy band. Since the on-axis X-ray sources can have PSF sizes under the ACIS pixel scale ($0.5\arcsec$), over-binning the images can better resolve the crowded core, while applying separate energy filters can potentially decompose individual sources that are otherwise blended in the broad-band image (e.g., a soft source in the vicinity of a hard source). 

\subsection{Source detection}
We use the {\sc ciao} {\tt wavdetect} script\footnote{\url{http://cxc.harvard.edu/ciao/ahelp/wavdetect.html}} \citep{Freeman02} to find and localise possible point sources in the field.  {\tt wavdetect} utilises a wavelet-based algorithm which correlates image pixels with the ``Mexican Hat" wavelet function at different scales. 
The tool searches for significant (at a given threshold) correlations and correspondingly centroids the sources, while calculating fluxes and other relevant properties. 

We set the {\tt scale} parameter to $1.0,~1.4,~2.0,~2.8,~\text{and}~4.0$ to account for sources of different sizes and use a significance threshold ({\tt sigthresh}) of $1.07\times 10^{-5}$ (reciprocal of the number of pixels in the image) to 
limit 
false detections. We first generate $3$ separate source lists by running {\tt wavdetect} on the extracted soft, hard and broad X-ray images (energy bands defined in Section \ref{sec:merge_chandra_obs}). These source lists are then cross-matched and concatenated to form a final source catalogue. 

Besides sources detected by \citet{Lugger07}, our {\tt wavdetect} run yields $9$ new sources ($>3~\sigma$) within the $1\farcm03$ half-light radius. To distinguish from previously detected sources, each new source is named as ``W" + a sequential number starting from 14. Positional and basic X-ray properties of all old and new sources are summarised in Table \ref{tab:x_ray_catalog}, and in Figure \ref{fig:x_ray_img} we present an X-ray image to show the spatial distribution of these sources.

The known MSP (PSR J2140--2310A; MSP A hereafter), and A3, a previously reported X-ray source, were not detected with the above {\tt wavdetect} parameters. The former has a radio position measured by timing observations as reported by \citet{Ransom04}. The latter is a faint source in close proximity to A1, which was reported as a detection by \citet{Lugger07}. These sources were detected with somewhat higher {\tt sigthresh} values, which might result in more spurious detections elsewhere but still provides good localisation. We found that both MSP A and A3 were detected when {\tt sigthresh}$=0.001$, with which {\tt wavdetect} found MSP A and A3 at the $2.9$ and $2.6\sigma$ level, respectively. Including MSP A, our X-ray catalogue has 10 new sources, extending the previous catalogue to 23 sources.

The updated half-light radius ($1\farcm03$) from \citet[2010 edition]{harris1996} is smaller than the $1\farcm15$ search radius 
used by \citet{Lugger07}, by which sources $12$ and $13$ are excluded. However, since these two sources were observed both in Cycle 3 and Cycle 18 and were not optically identified, we still include them in our analyses. 

\subsection{Source counts}
To calculate total source counts and fluxes, we use the {\sc ciao} {\tt srcflux} script. The script cannot be run on merged observations, so we applied the script on individual event files and then summed up the counts and averaged the fluxes. For each source, we compute source counts in the above-defined soft and hard bands, using circular extraction regions with radii that enclose roughly $90\%$ of the PSF in the broad band image. For each isolated source, we use the default setting where the background is defined by an annulus with inner radius equaling the source radius, and outer radius $5$ times that of the inner radius. Sources A1, A2, A3, MSP A, W17, C and W15 are in the close vicinity of another source (Figure \ref{fig:x_ray_img}). For these sources, we specify background regions that are outside the core and enclose only source-free fields. 

The resulting files from {\tt srcflux} include background-subtracted count rates and model-independent fluxes for individual observations. We convert the former to counts by multiplying by the corresponding exposure time, while the latter were converted to exposure-weighted fluxes. X-ray properties of all sources are summarised in Table \ref{tab:x_ray_catalog}.

\begin{table*}
\centering
\caption{M30 X-ray source catalogue.}
\renewcommand{\arraystretch}{1.2}
\begin{tabular}{ccccccccc}
\toprule
ID & $\alpha$ (ICRS) & $\delta$ (ICRS) & $P_\mathrm{err}^{a}$ & Offset$^b$ & \multicolumn{2}{c}{Counts$^c$}  & Flux ($0.5$--$7~\kev$)$^d$ & Type$^e$ \\
   &    (hh:mm:ss)    & ($^\circ$:$\arcmin$:$\arcsec$) & ($\arcmin$) & ($\arcsec$) & $0.5$--$2~\kev$ & $2$--$7~\kev$ & ($\times 10^{-16}~\mathrm{erg~s^{-1}~cm^{-2}}$) & \\
\midrule
A1    & 21:40:22.161   & $-$23:10:46.05   & $0.29$ & $0.03$ & $2452.3^{+81.7}_{-81.2}$ & $90.9^{+18.7}_{-13.5}$ & $540.0^{+19.3}_{-18.6}$ & qLMXB \\
A2    & 21:40:22.213   & $-$23:10:47.67   & $0.32$ & $0.03$ & $216.3^{+26.5}_{-22.4}$ & $42.0^{+13.2}_{-9.0}$ & $68.0^{+9.2}_{-7.0}$ & {\bf RS CVn?} \\
A3    & 21:40:22.026   & $-$23:10:47.62   & $0.42$ & $0.02$ & $45.6^{+14.3}_{-9.1}$ & $10.3^{+8.5}_{-3.8}$ & $22.2^{+8.0}_{-4.9}$ & {\bf CV?} \\
B     & 21:40:22.181   & $-$23:10:52.20   & $0.30$ & $0.08$ & $300.9^{+30.7}_{-27.3}$ & $199.2^{+25.8}_{-21.4}$ & $153.0^{+14.0}_{-11.7}$ & CV \\
C     & 21:40:22.954   & $-$23:10:49.75   & $0.30$ & $0.20$ & $433.6^{+35.6}_{-33.2}$ & $314.4^{+31.1}_{-27.6}$ & $232.0^{+17.1}_{-15.2}$ & CV \\
6     & 21:40:21.506   & $-$23:10:55.13   & $0.41$ & $0.19$ & $10.5^{+7.7}_{-4.0}$ & $6.1^{+6.7}_{-2.8}$ & $4.2^{+3.8}_{-1.7}$ & Unknown \\
7     & 21:40:21.598   & $-$23:10:33.14   & $0.47$ & $0.27$ & $9.7^{+7.4}_{-3.8}$ & $1.6^{+4.1}_{-1.2}$ & $6.0^{+9.7}_{-2.9}$ &  AB \\
8     & 21:40:22.123   & $-$23:11:14.48   & $0.35$ & $0.45$ & $33.8^{+12.7}_{-7.6}$ & $15.9^{+9.4}_{-5.0}$ & $14.3^{+5.3}_{-3.0}$ & {\bf CV?} \\
9     & 21:40:20.438   & $-$23:10:22.95   & $0.42$ & $0.56$ & $3.6^{+4.9}_{-2.2}$ & $1.9^{+4.1}_{-1.3}$ & $1.5^{+2.6}_{-0.8}$ & {\bf CV?} \\
10    & 21:40:23.237   & $-$23:09:59.25   & $0.53$ & $0.85$ & $5.8^{+5.6}_{-3.0}$ & $1.5^{+4.2}_{-1.1}$ & $1.0^{+1.2}_{-0.4}$ & AB \\
11    & 21:40:19.384   & $-$23:11:24.71   & $0.41$ & $0.88$ & $6.5^{+6.7}_{-2.9}$ & $10.2^{+8.3}_{-3.7}$ & $8.3^{+7.1}_{-2.9}$ & Unknown \\
12    & 21:40:27.029   & $-$23:10:39.24   & $1.09$ & $1.14$ & $18.9^{+10.8}_{-5.2}$ & $14.2^{+9.4}_{-4.4}$ & $13.4^{+7.1}_{-3.5}$ & {\bf CV?} \\
13    & 21:40:26.569   & $-$23:11:18.81   & $0.97$ & $1.15$ & $17.0^{+9.9}_{-5.1}$ & $65.1^{+16.4}_{-11.2}$ & $49.1^{+11.8}_{-8.1}$ & {\bf AGN?} \\
MSP A & 21:40:22.403   & $-$23:10:48.68   & $0.43$ & $0.07$ & $21.7^{+10.9}_{-5.9}$ & $3.3^{+5.8}_{-1.6}$ & $6.7^{+4.4}_{-1.8}$ & {\bf MSP} \\
W14   & 21:40:25.505   & $-$23:11:18.55   & $0.37$ & $0.94$ & $5.2^{+6.8}_{-2.3}$ & $25.7^{+11.6}_{-6.4}$ & $15.1^{+6.6}_{-3.7}$ & Unknown \\
W15   & 21:40:22.833   & $-$23:10:47.50   & $0.43$ & $0.17$ & $4.0^{+6.0}_{-2.1}$ & $9.5^{+8.0}_{-3.6}$ & $5.3^{+4.8}_{-1.9}$ & {\bf CV} \\
W16   & 21:40:20.963   & $-$23:10:43.51   & $0.44$ & $0.27$ & $10.1^{+8.0}_{-3.8}$ & $7.3^{+6.6}_{-3.3}$ & $9.6^{+6.8}_{-3.4}$ & {\bf RS CVn?} \\
W17   & 21:40:22.182   & $-$23:10:43.53   & $0.46$ & $0.07$ & $17.6^{+10.3}_{-5.0}$ & $10.1^{+8.3}_{-3.6}$ & $12.7^{+6.9}_{-3.4}$ & {\bf AB?} \\
W18   & 21:40:18.213   & $-$23:10:39.10   & $0.59$ & $0.91$ & $1.6^{+4.1}_{-1.1}$ & $3.8^{+6.1}_{-2.0}$ & $1.5^{+2.4}_{-0.8}$ & Unknown \\
W19   & 21:40:23.932   & $-$23:10:13.85   & $0.48$ & $0.70$ & $11.1^{+8.5}_{-3.9}$ & $6.0^{+6.7}_{-2.7}$ & $4.9^{+3.7}_{-1.6}$ & {\bf CV?} \\
W20   & 21:40:24.098   & $-$23:11:33.31   & $0.78$ & $0.89$ & $2.8^{+4.0}_{-2.1}$ & $3.4^{+5.8}_{-1.7}$ & $2.1^{+3.0}_{-0.9}$ & {\bf AB?} \\
W21   & 21:40:25.035   & $-$23:10:35.59   & $0.53$ & $0.70$ & $8.3^{+7.5}_{-3.3}$ & $0.7^{+3.0}_{-0.7}$ & $3.6^{+6.4}_{-2.0}$ & {\bf CV} \\
W22   & 21:40:22.234   & $-$23:09:50.95   & $0.70$ & $0.94$ & $<2.3$ & $5.4^{+6.4}_{-2.5}$ & $4.5^{+5.8}_{-2.3}$ & Unknown \\
\bottomrule
\multicolumn{8}{l}{$^a$95\% error radii calculated according to \citet{hong2005}.}\\
\multicolumn{8}{l}{$^b$Offsets from the cluster centre.}\\
\multicolumn{8}{l}{$^c$Source counts as calculated by {\tt srcflux}; the errors are at the 90\% confidence level.}\\
\multicolumn{8}{l}{The 90\% upper limits are calculated according to \citet{gehrels1986}}\\
\multicolumn{8}{l}{$^d$Model-independent fluxes as calculated by {\tt srcflux}; the errors are at the 90\% confidence level.}\\
\multicolumn{8}{l}{$^e$Bold texts indicate new classifications compared to \citet{Lugger07}.}
\end{tabular}
\label{tab:x_ray_catalog}
\end{table*}

\begin{figure*}
    \centering
    \includegraphics[scale=0.35]{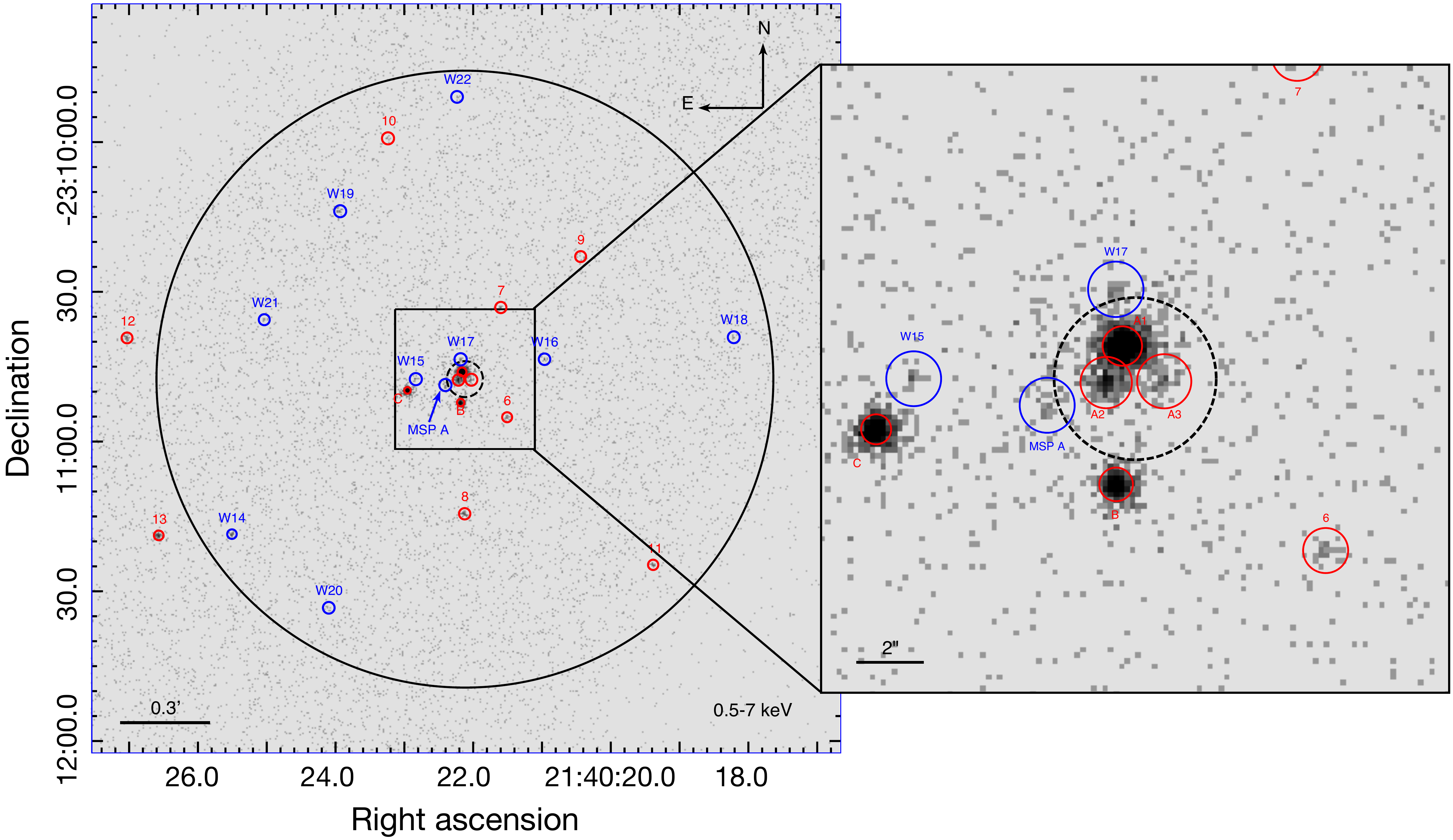}
    \caption{$0.5$--$7~\kev$ X-ray image of M30. The left panel shows a $2\farcm5\times 2\farcm5$ square region centered on the cluster. All sources are indicated with circles that enclose $90\%$ of their PSFs, with new sources in blue and previously reported sources \citep{Lugger07} in red. The solid black circle indicates the $1\farcm 03$ half-light radius of the cluster and the dashed black circle represents the $0\farcm06$ core region according to \citet[2010 edition]{harris1996}. The right panel shows a zoomed-in view of the central $28\arcsec \times 28\arcsec$ square region.}
    \label{fig:x_ray_img}
\end{figure*}

\subsection{X-ray spectral analyses}
Spectral analyses of the qLMXB A1 to constrain the NS radius and mass are presented in a separate work \citep{Echiburu20}. The analyses in this paper focus on the other, especially the newly detected, sources, using mostly the Cycle 18 data. 

We first extract X-ray spectra using the {\sc ciao} {\tt specextract} tool\footnote{\url{http://cxc.harvard.edu/ciao/ahelp/specextract.html}}. The source regions are defined as circular regions that enclose roughly $90\%$ of the source PSF, while background regions are chosen as source-free annulus regions around (in cases of sources not affected by crowding) or away from the source (in cases of sources close to the crowded core).

Spectral analyses were performed with the {\sc heasoft/xspec} software (version 12.10.1; \citealt{Arnaud96}). We combined the Cycle 18 spectra and the corresponding associated files (including response matrices, ancillary response files, and background spectra), using the {\tt addspec} script in {\sc heasoft/ftools}, and properly rebin the co-added spectra using the {\sc ciao} {\tt dmgroup} tool. For sources A2, B and C, which have more than $100$ counts, we rebinned the spectra to at least 10 counts per bin and analyse them using $\chi^2$ statistics; whereas for sources with fewer than $100$ counts, we binned them to contain at least $1$ count per bin and use W-statistics \citep{cash1979} in our analyses, while applying the Cramer von Mises (cvm) test statistics. \cory{We found that source 10 has too few Cycle 18 counts ($3$ counts between $0.5$--$10~\mathrm{keV}$) to properly constrain the spectral flux, so we also incorporated the Cycle 3 data.} The fitting quality is therefore evaluated either through the reduced $\chi^2$ ($\chi_\nu^2=\chi^2/\mathrm{dof}$) or, for W-statistics, roughly through the {\tt goodness} command in {\sc xspec}, which generates a given number of simulated spectra based on the best-fitting parameters and calculates the fraction of realisations with the fit statistic lower than that for the data. For each W-statistics fit, we generate 1000 realisations and expect percentages below $90\%$ as acceptable. All parameters are reported at the $90\%$ confidence level.


%
We tried to fit the rebinned spectra to an absorbed power-law ({\tt pow}) or thermal plasma model ({\tt vapec} in {\sc xspec}) between $0.5$ and $10~\mathrm{keV}$, keeping models that better describe the spectra. The Fe abundance in {\tt vapec} is fixed at the cluster value ($\mathrm{[Fe/H]}=-2.27$ from \citealt{harris1996}). Galactic absorption is accounted for by convolving the models with the {\tt tbabs} model in {\sc xspec} using the {\tt wilms} abundance \citep{Wilms00}. We keep the hydrogen column density ($N_\mathrm{H}$) as a free parameter for sources with more than $100$ counts, while the $N_\mathrm{H}$ for fainter sources is fixed at the cluster value ($\approx 2.61\times 10^{20}~\mathrm{cm^{-2}}$, calculated using the cluster reddening $E(B-V)=0.03$ from \citealt{harris1996} and a conversion factor from \citealt{Bahramian15}). For the faintest spectra that have less than $10$ counts (source 10, W17, W18 and W20), constraints on model parameters derived directly from the fits become tenuous, so we only fit the normalisation parameters for these spectra. The plasma temperature (for the {\tt vapec} model) or photon index ($\Gamma$ in the {\tt pow} model) are fixed to averaged values obtained by fitting a {\tt vapec} or {\tt pow} to the co-added spectra of all the faint source spectra ($\Gamma=2$ for {\tt pow}; $kT=2.7~\mathrm{keV}$ for {\tt vapec}). For the known MSP, MSP A, we fit both a {\tt pow} model and a blackbody model, {\tt bbodyrad}, to its spectrum. We report our results in Table \ref{tab:spectral_fits}. 


\begin{table*}
    \centering
    \caption{Results of spectral fitting to the Cycle 18 spectra.}
    \renewcommand{\arraystretch}{1.2}
    \begin{tabular}{cccccccc}
    \toprule
    ID  &  Model  &  $N_\mathrm{H}$ & $kT^a$ & $\Gamma$ or $R_\mathrm{bb}^b$ & $F_X$($0.5$--$2~\mathrm{keV}$)$^c$ & $F_X$($2$--$7~\mathrm{keV}$)$^d$ & $\chi_\nu^2 (\mathrm{dof})$ or Goodness\\
    \midrule
        &  {\sc xspec} & $10^{20}~\mathrm{cm^{-2}}$ & $\mathrm{keV}$ &  & $10^{-16}~\mathrm{erg~s^{-1}~cm^{-2}}$ &
        $10^{-16}~\mathrm{erg~s^{-1}~cm^{-2}}$ & \\ 
    \midrule
    A2  & {\tt vapec} & $<8.0^\ast$ & $4.1^{+4.4}_{-2.2}$ & - & $21.0^{+3.1}_{-3.1}$ & $23.4^{+3.5}_{-3.5}$ & $1.20~(11)$ \\
    A3  & {\tt vapec} & $2.6^\dag$  & $2.0^{+2.9}_{-1.0}$ & - & $6.0^{+2.1}_{-1.7}$ & $3.7^{+1.3}_{-1.0}$ & $65.40\%$ \\
    B   & {\tt vapec} & $<12^\ast$  & $10.2^{+9.8}_{-4.6}$ & - & $64.2^{+5.1}_{-5.1}$ & $109.6^{+8.6}_{-8.6}$ & $0.98~(44)$ \\
    C   & {\tt vapec} & $18.9^{+10.5}_{-8.5}$ & $13.2^{+13.1}_{-5.2}$ & - & $130.7^{+7.5}_{-7.5}$ & $243.7^{+13.9}_{-13.9}$ & $0.85~(74)$ \\
    6   & {\tt vapec} & $2.6^\dag$ & $>2.6$ & - & $2.6^{+1.4}_{-1.1}$ & $5.7^{+3.1}_{-2.4}$ & $9.30\%$ \\
    7   & {\tt vapec} & $2.6^\dag$  & $0.4^{+0.5}_{-0.2}$ & - & $4.0^{+3.7}_{-2.3}$ & $0.04^{+0.04}_{-0.03}$ & $64.40\%$ \\
    8   & {\tt vapec} & $2.6^\dag$  & $3.4^{+11.2}_{-1.7}$ & - & $7.3^{+2.2}_{-1.9}$ & $7.2^{+2.2}_{-1.8}$ & $40.50\%$ \\
    9   & {\tt vapec} & $2.6^\dag$  & $0.7^{+44.8}_{-0.4}$ & - & $1.9^{+1.8}_{-1.2}$ & $0.2^{+0.2}_{-0.1}$ & $37.30\%$ \\
    10  & {\tt pow}   & $2.6^\dag$  & - & $2^\dag$ & $2.0^{+1.6}_{-1.1}$ & $1.8^{+1.5}_{-1.0}$ & $25.60\%$ \\ 
    11  & {\tt pow}   & $2.6^\dag$ & - & $-0.2^{+0.9}_{-1.0}$ & $1.1^{+0.6}_{-0.4}$ & $16.4^{+8.9}_{-6.7}$ & $29.00\%$ \\
    12  & {\tt vapec} & $2.6^\dag$ & $>3.4^\ast$ & - & $5.8^{+1.9}_{-1.6}$ & $10.1^{+3.3}_{-2.8}$ & $65.10\%$ \\
    13  & {\tt pow} & $2.6^\dag$ & - & $0.0^{+0.3}_{-0.4}$ & $4.4^{+0.9}_{-0.8}$ & $51.0^{+10.0}_{-8.9}$ & $43.00\%$ \\
    MSP A & {\tt bbodyrad} & $2.6^\dag$ & $0.3^{+0.1}_{-0.1}$ & $70^\ast$ & $4.9^{+2.1}_{-1.7}$ & $0.6^{+0.2}_{-0.2}$ & $27.00\%$ \\
    MSP A & {\tt pow} & $2.6^\dag$ & - & $2.9^{+0.9}_{-0.9}$ & $6.2^{+2.6}_{-2.1}$ & $1.8^{+0.7}_{-0.6}$ & $72.60\%$ \\
    W14  & {\tt pow} & $2.6^\dag$ & - & $0.2^{+0.6}_{-0.6}$ & $1.9^{+0.6}_{-0.5}$ & $18.6^{+6.5}_{-5.3}$ & $51.80\%$ \\
    W15  & {\tt vapec} & $2.6^\dag$ & $>6.6^\ast$ & - & $2.2^{+1.2}_{-0.9}$ & $5.3^{+2.8}_{-2.1}$ & $40.30\%$ \\
    W16 & {\tt vapec} & $2.6^\dag$ & $>1.0^\ast$ & - & $4.3^{+1.9}_{-1.5}$ & $3.9^{+1.7}_{-1.4}$ & $49.30\%$ \\
    W17 & {\tt pow}   & $2.6^\dag$ & - & $1.1^{+0.9}_{-0.9}$ & $1.6^{+0.9}_{-0.7}$ & $4.8^{+2.8}_{-2.1}$ & $64.40\%$ \\
    W18 & {\tt pow}   & $2.6^\dag$ & - & $2^\dag$ & $0.9^{+1.4}_{-0.8}$ & $2.5^{+2.9}_{-1.7}$ & $88.40\%$ \\
    W19 & {\tt vapec} & $2.6^\dag$ & $>2.0^\ast$ & - & $2.3^{+1.1}_{-0.9}$ & $5.2^{+2.6}_{-2.0}$ & $55.90\%$ \\
    W20 & {\tt pow}   & $2.6^\dag$ & - & $2^\dag$ & $1.8^{+2.2}_{-1.2}$ & $1.4^{+2.4}_{-1.2}$ & $52.80\%$ \\
    W21 & {\tt vapec} & $2.6^\dag$ & $0.6^{+3.9}_{-0.3}$ & - & $2.4^{+2.1}_{-1.1}$ & $0.2^{+0.1}_{-0.1}$ & $10.50\%$ \\
    W22 & {\tt pow}   & $2.6^\dag$ & - & $-3.0^{+1.9}_{-2.6}$ & $0.02^{+0.02}_{-0.01}$ & $11.7^{+8.8}_{-6.1}$ & $67.00\%$ \\
    \bottomrule
    \multicolumn{8}{l}{$^a$Plasma temperature (of the {\tt vapec}) model or blackbody temperature (of the {\tt bbodyrad}) model.}\\
    \multicolumn{8}{l}{$^b$ $R_\mathrm{bb}$ is the radius of the emission region (in m) derived from the {\tt bbodyrad} model.}\\
    \multicolumn{8}{l}{$^{c,d}$Unabsorbed model fluxes.}\\
    \multicolumn{8}{l}{$^\ast$ superscripts indicate that either the upper or lower limit (or both) of the parameter is (are) unconstrained.}\\
    \multicolumn{8}{l}{$^\dag$ superscripts mark parameters fixed during the fit.}
    \end{tabular}
    \label{tab:spectral_fits}
\end{table*}

\subsection{X-ray variability}
We run the {\sc ciao} {\tt glvary} script to check for variability within each {\it Chandra} exposure following the instructions in the ``Searching for Variability in a Source" thread\footnote{\url{https://cxc.cfa.harvard.edu/ciao/threads/variable/}}. The {\tt glvary} script applies the Gregory-Loredo algorithm \citep{Gregory92}, which divides the event time series into multiple time bins and looks for significant variations between these bins. The degree of variability is indicated with a variability index (VI) between 0 and 10, with values $\geq 6$ considered confident variables. 

We found variability for A2 and W16 (VI=6) in obsID. 20792 and 18997 (Table \ref{tab:x_ray_obs}), respectively; likely variability (VI=5) was found in source 8 in obsID. 18997. As expected, we did not find variability for A1 (VI=0) in any of the observations, as emission from qLMXBs are dominated by non-varying thermal X-rays originating from the NS surface.

\subsection{Astrometry}
\subsubsection{
Co-adding images
}
We used the Python routine {\tt astrodrizzle} from the {\sc drizzlepac} software package\footnote{\url{http://www.stsci.edu/scientific-community/software/drizzlepac.html}} (version 2.0) to generate combined optical images. The {\tt astrodrizzle} routine takes advantage of the dithering information from the single FLC frames to remove cosmic rays and small-scale detector defects (e.g., hot pixels, bad columns, etc.), while generating co-added images and reconstructing information lost due to undersampling.

For each filter, the individual FLC frames are first aligned with the {\tt Tweakreg} tool, accounting for shifts between the FLC frames. The aligned FLC frames are then combined using {\tt astrodrizzle}. We set the {\tt pixfrac} parameter to 1.0 and use a final pixel size ({\tt final\_scale}) half of the native pixel scale ($0.02\arcsec$ for WFC3 and $0.025\arcsec$ for ACS), oversampling the co-added image by a factor of $2$ to mitigate crowding in the core. The ``drizzle"-combined images are then used for further astrometric analyses.

\subsubsection{Absolute Astrometry}
There are multiple factors that may affect the accuracy of the {\it HST} astrometry solution, but the main source of error comes from uncertainties in guide star positions, from which astrometry is derived. As the uncertainties are typically $\approx 200~\mathrm{mas}$ and $\approx300~\mathrm{mas}$ for WFC3 and ACS, respectively \citep{lucas2016, deustua2016}, they might significantly alter our 
counterpart identifications\footnote{At the time of writing, astrometry of WFC3 and ACS imaging products retrieved from the STSCI archive have been aligned to an updated guide star catalogue, and some have been directly aligned to Gaia catalogue. More details can be found in \url{https://archive.stsci.edu/contents/newsletters/december-2019/improved-astrometry-for-wfc3-and-acs?filterName=news-filter&filterPage=news}}.

To correct our absolute astrometry, we aligned the {\it HST} images to the {\it Gaia} DR2 catalogue \citep{gaia2016a, gaia2018}, which offers 
accurate source positions with uncertainties on milliarcsecond ($\mathrm{mas}$) scales. Since most well-localised {\it Gaia} sources in M30 are bright, we chose the $\U$ image as the reference frame that is sufficiently long in exposure to include  many  bright objects, but not too deep so as to saturate these stars. In fact, the drizzle-combined dithered images  improve spatial sampling of the PSF and therefore provide somewhat improved centroiding.

We found 217 {\it Gaia} sources within the half-light radius that have superior astrometric solutions (uncertainties in both RA and DEC $\leq 0.2~\mathrm{mas}$). 
On the $\U$ image, we obtained centroid positions of stars by running the {\tt daofind} task in the {\sc daophot} software package \citep{stetson1987}. In the next step, we cross-match the {\it Gaia} catalogue with the $\U$ image to find counterparts to the {\it Gaia} sources, finding $217$ matches with average offsets in RA and DEC ({\it Gaia}$-${\it HST}) $=0.135\arcsec \pm 0.006\arcsec$ and $0.123\arcsec \pm 0.004\arcsec$ ($1~\sigma$ errors). Finally, the offsets were applied to update the WCS information of the $\U$ image, to which the other {\it HST} images are aligned.

\subsubsection{{\it Chandra} boresight correction}
We have applied relative shifts to individual ACIS observations prior to merging (Section \ref{sec:merge_chandra_obs}); however, the resulting merged event file still requires further  astrometric corrections before proceeding with counterpart searches. 

For this purpose, we chose the on-axis sources B and C which have relatively well-defined PSFs and possess well-identified counterparts \citep{Lugger07}. We calculated the corresponding shifts between {\tt wavdetect}-determined X-ray positions and the corresponding counterpart {\tt daofind} positions, yielding average shifts ({\it Gaia $-$ Chandra}) $\approx -0.279\arcsec$ and $\approx 0.107\arcsec$ in RA and DEC, respectively, which were then applied to the WCS header information to correct the boresight.

\subsection{Optical photometry}
\label{sec:optical_photometry}
The photometry of most stars has been fully analysed as a part of the HUGS project \citep{piotto2015}. The catalogue data products are now available to the public \citep{Nardiello18}, from which we can readily construct colour-magnitude diagrams (CMDs). However, we noted that some stars (especially the faint ones), though clearly present in the field, do not have HUGS photometry. We therefore also generate our own photometric catalogue as a complement to the HUGS results. 

We use the {\sc dolphot} (version 2.0) photometry package (software based on {\sc hstphot}; \citealt{Dolphin2000}) to generate photometry catalogues for individual frames observed by WFC3 and ACS. First, for each camera, we use a drizzle-combined image ($\U$ for WFC3; $\I$ for ACS)\footnote{Note that the drizzle-combined images used here are on the original pixel scales, i.e., not oversampled.} as the reference frame, to which coordinates of stars found on individual exposures are transformed. We then applied the {\tt wfc3mask} and the {\tt acsmask} tools to the individual WFC3 and ACS exposures to mask the flagged bad pixels while multiplying the pixel area map by each FLC image. In the next step, we use the {\tt splitgroup} tool to extract 2 individual science extensions from each FLC image, corresponding to separate exposures on the two CCD chips. These CCD-specific images are then fed to the {\tt calcsky} tool to calculate sky levels, generating a sky map for each image. 

The above processes set up the basic input to the final runs. The final step prior to running the {\tt dolphot} task is to determine the proper parameters. We list in Table \ref{tab:dol_parameter} the parameters we used that are not at their default values. We run {\tt dolphot} using the Anderson PSF library (WFC3 ISR 2018-14; ACS ISR 2006-01) by specifying the {\tt ACSpsfType} and {\tt WFC3UVISpsfType} parameters. The software then performs aperture and PSF photometry on the individual FLC frames, calibrates the photometry to big apertures ($0.6\arcsec$ for WFC3; $0.5\arcsec$ for ACS), and converts the magnitudes to the VEGMAG system using up-to-date zeropoints for ACS \citep{sirianni2005} and WFC3\footnote{\url{https://www.stsci.edu/hst/instrumentation/wfc3/data-analysis/photometric-calibration/uvis-photometric-calibration}}. 

The resulting catalogues include the basic positional information and instrumental magnitudes for stars cross-identified on multiple filters. Besides, there are multiple quality parameters that can be used to significantly reduce the raw catalogues and cull for stars with high quality photometry. For each camera, we first extracted a sub-catalogue that contains all star-like objects that have valid instrumental magnitudes in at least one filter. This catalogue is further used for identification of counterparts. We also extract sub-catalogues for making $\UV-\U$, $\U-\B$ and $\V-\I$ colour-magnitude diagrams (CMDs; Figure \ref{fig:uv_and_vi_cmds}). For this purpose, we use stars with relatively low ($<0.1$) uncertainties in photometry, which significantly reduced the spread in the faint end of the CMDs. Note that this reduction is only to improve readability of the plot; for faint counterparts that have intrinsically high uncertainties, we keep the error bars when we plot them on the CMDs. 

Our photometry catalogue might complement the HUGS photometry in the WFC3 bands ($\UV$, $\U$, and $\B$). However, in the ACS images ($\V$ and $\I$), magnitudes of some stars were not properly measured by {\sc dolphot}. Particularly in the core, photometry of some stars is not available due to crowding or bleeding patterns from bright stars, albeit the target is clearly detected. The HUGS results might also miss some stars but are superior to our photometry in $\V$ and $\I$. Therefore, for stars not measured by {\sc dolphot}, we search in the HUGS catalogue and calibrate the magnitudes to align with our photometry (indicated with $^\dag$ superscripts in Table  \ref{tab:opt_counterparts}). This conversion was done by calculating average offsets, in $\V$ and $\I$, using stars with relatively small ($\leq 0.05$ in {\sc dolphot} and $\leq 0.01$ in HUGS) photometric uncertainties. We found 12086 and 12540 such matches in $\V$ and $\I$ bands and obtained 3 $\sigma$-clipped median offsets ({\sc dolphot}-HUGS) of $-0.013$ and $0.029$ (with 1 $\sigma$ error of $\approx 0.01$ in both bands), respectively.

\subsubsection{UV variability}
The {\sc dolphot} output also includes frame-specific magnitudes that allow us to check the variability of interesting targets. For this purpose, we collect magnitudes measured on multiple frames and plot their RMS values ($\sigma$'s) vs. mean magnitudes. The results are presented in Figure \ref{fig:uv_variability}. While most stars lies on a sequence with increasing RMS towards fainter magnitudes, stars with strong variability should be outliers from the sequence. 

\begin{table}
    \centering
    \caption{Parameters used in {\sc dolphot}.}
    \resizebox{\columnwidth}{!}{
    \begin{tabular}{cccl}
    \toprule
       Name  & \multicolumn{2}{c}{Value} & Notes \\
             & WFC3 & ACS & \\
    \midrule
    {\tt img\_RAper} & $8$ & $8$ & \makecell[l]{Photometry apertures} \\ [0.3cm]
    {\tt img\_RSky}  & $9,14$ & $9,14$ &\makecell[l]{ Radii of sky annulus} \\ [0.3cm]
    {\tt img\_RPSF}  & $15$ & $15$ & PSF radius \\ [0.3cm]
    {\tt img\_aprad} & $15$ & $10$ & \makecell[l]{Radius for aperture \\ correction} \\[0.3cm]
    {\tt SigFind}    & $3$  & $3$  & Detection threshold ($\sigma$) \\[0.3cm]
    {\tt UseWCS}     & $1$  & $1$  & Use WCS in alignment \\[0.3cm] 
    {\tt Align}      & $1$  & $1$  & Offsets and re-scaling \\ [0.3cm]
    {\tt ACSpsfType} & $0$  & $1$  & \makecell[l]{Use Anderson PSF \\ cores for ACS}\\ [0.3cm]
    {\tt WFC3UVISpsfType} & $1$ & $0$ & \makecell[l]{Use Anderson PSF \\ cores for WFC3}\\ [0.3cm]
    \bottomrule
    \end{tabular}
    }
    \label{tab:dol_parameter}
\end{table}

\subsection{Counterpart search and source identification}
\label{sec:counterpart_search}
\subsubsection{Basic methodologies}
Identifying X-ray sources involves incorporation of positional, photometric (both optical and X-ray), temporal, and membership information of each source. Positionally, since bright ($\gtrsim 100$ counts) X-ray sources are more accurately localised, we primarily search within their $95\%$ error circles (Table \ref{tab:x_ray_catalog}), expecting optical counterparts close to the nominal X-ray positions. Faint X-ray sources might have more uncertain X-ray positions, we thus expand the searching region to $1.5$ times their $95\%$ error radii. For sources that have {\it VLA} counterparts and thus somewhat better localisation, we search the radio error regions ($\sim 1/10$ the synthesised beam) for optical counterparts.

The photometric criteria used for identifying optical counterparts depend on the type of the source. For example, we identify CVs with blue outliers on the UV CMDs, as CVs usually manifest strong UV emission thought to originate from a shock-heated region on the white dwarf surface (in the case of magnetic CVs) and/or from the accretion disc; whereas, when observed in optical bands like $\V$ and $\I$, optical fluxes are dominated by the companion, which has a lower surface temperature, so they may appear to be consistent with, or even slightly redder than, the main sequence \citep[e.g.,][]{Edmonds03a, Edmonds03b, Lugger17, Zhao19}. On the other hand, ABs often manifest slight red excesses 
consistent with the binary sequence (BY Dra systems), or they have subgiant or red giant companions (RS CVn systems). 

Care should be taken, however, that background AGN can masquerade as CVs with blue excesses on multiple CMDs; whereas foreground stars can be strong X-ray emitters and red outliers. The AGN nature of some sources might be indicated by detections of radio counterparts, yet for most sources that have no radio counterparts, we may rely on the kinematic information of the optical counterpart to check for membership. In this regard, cluster membership probabilities derived from proper motions (PMs) from the HUGS catalogue \citep{Nardiello18} are very useful in our analyses. 


In X-ray, we define a hardness ratio
\begin{equation}
    X_C = 2.5\log_{10}\left(\dfrac{F_{0.5-2}}{F_{2-7}}\right),
    \label{eq:x_ray_color}
\end{equation}
where $F_{0.5-2}$ and $F_{2-7}$ are the unabsorbed model fluxes in the soft and hard bands from our spectral fits (Table \ref{tab:spectral_fits}), which can roughly separate different source classes. In Figure \ref{fig:x_ray_cmd}, we present an X-ray colour-magnitude diagram, plotting $0.5$--$7~\mathrm{keV}$ fluxes (and luminosities calculated assuming the cluster distance) vs. $X_C$. 
Comparison with e.g. the X-ray hardness diagram of \citet{heinke2005} shows that consistency with a photon index around 1.7 is common for bright ($L_X>10^{31}~\mathrm{erg~s^{-1}}$) CVs (and, more rarely, bright MSPs and ABs), as well as AGN. CVs and AGN can also show harder spectra, if they show internal absorption (e.g., they are seen edge-on through their accretion discs). Consistency with a photon index between 2 and 4 is typical of fainter ($L_X<10^{31}~\mathrm{erg~s^{-1}}$) MSPs, CVs, and (very common) ABs, and of quiescent LMXBs at any luminosity. 


Source classification can be further complemented by investigation of X-ray/optical ratios. In Figure \ref{fig:lxmv}, we plot $0.5$--$2.5~\mathrm{keV}$ X-ray luminosities vs. absolute $V$ band magnitudes ($M_V$) for sources identified with optical counterparts. \citet{Bassa04} found an empirical separatrix on this diagram that roughly distinguishes cluster ABs from CVs, given by $\log_{10}L_X[0.5-2.5~\mathrm{keV}] = -0.4M_V + 34$. CVs usually have higher ratios and locate above this separatrix, while ABs lie to the lower right in the $L_X-M_V$ plane.

Finally, temporal features, especially signs of variability, might also be useful in determining the source nature.
Most classes of X-ray source show variability; the only classes that lack evidence of intrinsic variability (on timescales of minutes to days) are quiescent LMXBs and MSPs showing only thermal X-ray emission from their surfaces, or MSPs showing magnetospheric emission (note that redback MSPs showing shock-powered emission do vary). 

\subsubsection{Radio counterparts}
Radio counterparts to X-ray sources can provide complementary information on the source nature. Indeed, multiple source classes are also expected to be radio emitters. An important feature is the radio spectral index ($\alpha$; as defined in $S_\nu \propto \nu^{\alpha}$, where $S_\nu$ is the specific flux density). In both NS \citep{Migliari06} and BH LMXBs \citep{Gallo14, Gallo18}, radio emission has been observed to have flat ($-0.5<\alpha<0$) to slightly inverted ($\alpha >0$) spectra, which are thought to be from relativistic jets; whereas MSPs commonly exhibit steep radio spectra ($\alpha \approx -1.4$ with unit standard deviation; see \citealt{Bate13}). Radio emission has also been observed in CVs \citep[e.g.,][]{Coppejans15, Coppejans16, Barrett17}, which are usually fainter than LMXBs but could attain higher flux levels during occasional flares \citep[e.g.,][]{Russell16}. ABs can also be radio sources, but even short bright flares, as bright as $10^{19}~\mathrm{erg~s^{-1}~Hz^{-1}}$ at centimeter wavelengths \citep[see e.g.,][]{Drake89, Osten00}, are too faint to be detected by our {\it VLA} observation considering the distance to M30. Finally, supermassive black holes in AGN are also active radio emitters; \toreferee{indeed, AGNs are the most common radio sources observed in the directions of GCs (Shishkovsky, L et al., submitted to ApJ).}

\subsubsection{Chance coincidence}
\label{sec:chance_coincidence}
Optical counterparts that are positionally consistent with the X-ray error circles might be chance coincidences. This is significantly more likely in the dense cluster core. To estimate the average number of chance coincidences, we use the $\UV-\U$ CMD from the HUGS catalogue, plotted only for stars that are likely cluster members ($P_\mu\geq 90\%$). We then apply polygonal selection areas to roughly separate different subpopulations (Figure \ref{fig:subpopulations}) using the {\sc glueviz} software \citep{Beaumont15, Robitaille17}. There are a total of $21596$ cluster members detected in the WFC3 field of view (FOV; $160\arcsec\times 160\arcsec$), including $20256$ main sequence stars, $744$ evolved members, including $338$ subgiants and $406$ red giants, $163$ blue stars ($\UV -\U \lesssim 0.72$), $151$ red stars ($\UV-\U \gtrsim 0.58$), $49$ blue stragglers (BSs) and $12$ sub-subgiant stars (SSGs)\footnote{Stars that are lower than the subgiant branch but redder than the main sequence; see \citet{Leiner17}}.

To estimate numbers of chance coincidences for different subpopulations, we divide the cluster into a series of concentric annuli of the width of the core radius $0\farcm06$. Stars of different subpopulations are assumed to be evenly distributed within each annulus, so the number of chance coincidences is roughly

\begin{equation}
    N_\mathrm{c} = N_\mathrm{tot} \times \frac{\overline{A_\mathrm{err}}}{A_\mathrm{annu}},
    \label{eq:num_chance_coincidence}
\end{equation}
where $N_\mathrm{c}$ is the number of chance coincidences within a given annulus, $N_\mathrm{tot}$ is the total number of stars in one of the subpopulations, $\overline{A_\mathrm{err}}$ is the averaged area of the 95\% error region derived from the average error radius ($\approx 0.5\arcsec$) and $A_\mathrm{annu}$ is the area of the given annulus. BS and SSG populations have small numbers of stars, so we only make the estimate over the whole WFC3 FOV, which yields $N_\mathrm{c}\approx1.5\times 10^{-3}$ and $\approx 3.7\times 10^{-4}$ for BSs and SSGs, respectively. $N_\mathrm{c}$ values for other populations are presented in Figure \ref{fig:chance_coincidence}, and it is clear that $N_\mathrm{c}$ exhibits a general descending trend toward the cluster outskirts; specifically, $N_\mathrm{c}$ is roughly $10$ times higher in the core than at the half-light radius for main sequence stars.


\begin{figure}
    \centering
    \includegraphics[width=\columnwidth]{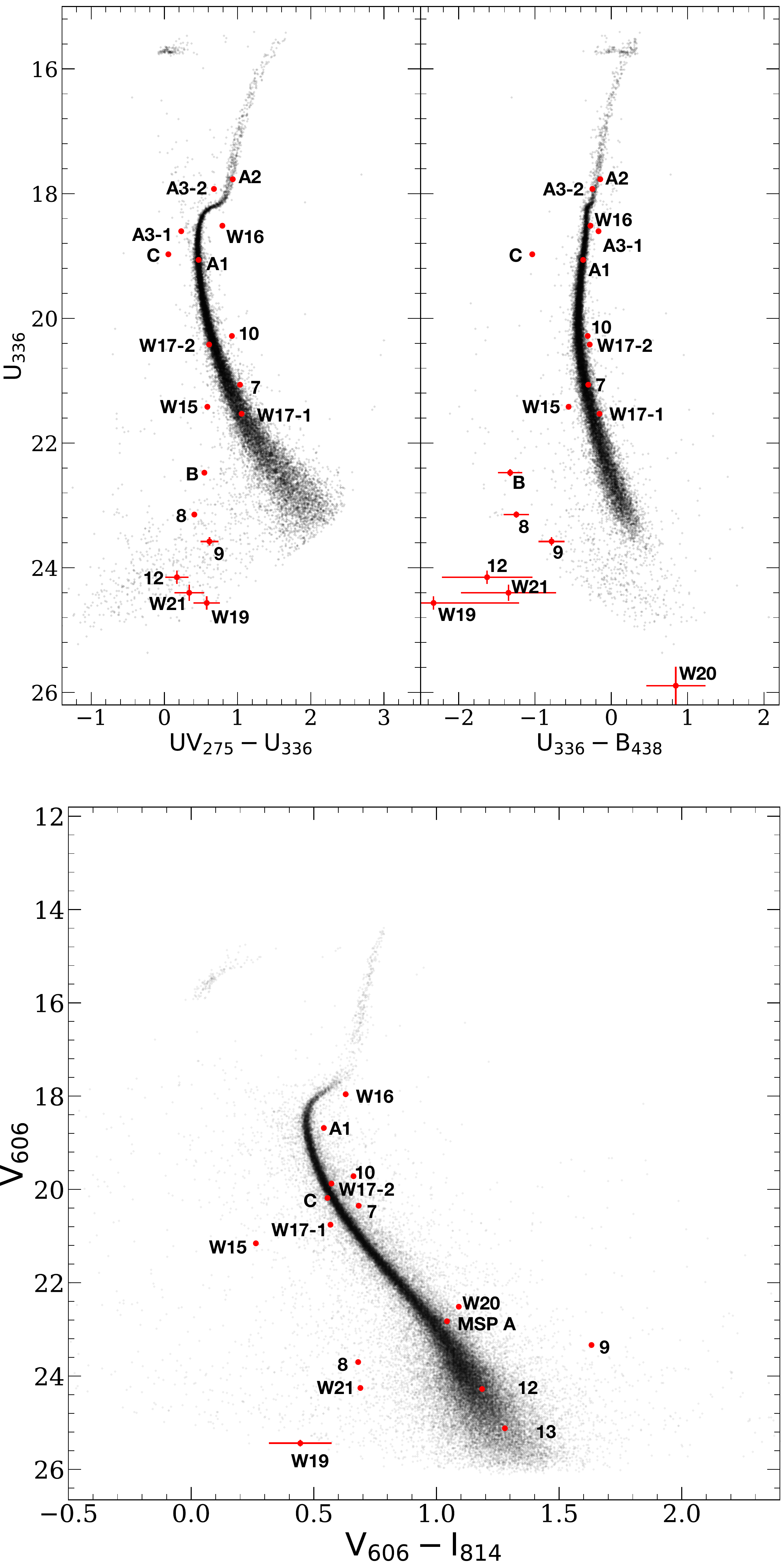}
    \caption{\toreferee{{\it Top}: $\UV-\U$ (top left) and $\U-\B$ (top right) CMDs plotted for stars with errors in magnitudes $\leq 0.1$ (black). {\it Bottom}: $\V-\I$ CMD plotted for stars with errors in magnitudes $\leq 0.1$ (black). The locations of the counterparts are marked with red dots. Counterparts with uncertainties in magnitudes greater than $0.1$ (intrinsic error calculated by {\sc dolphot}) are plotted together with error bars.}}
    \label{fig:uv_and_vi_cmds}
\end{figure}


\begin{figure*}
    \centering
    \includegraphics[scale=0.45]{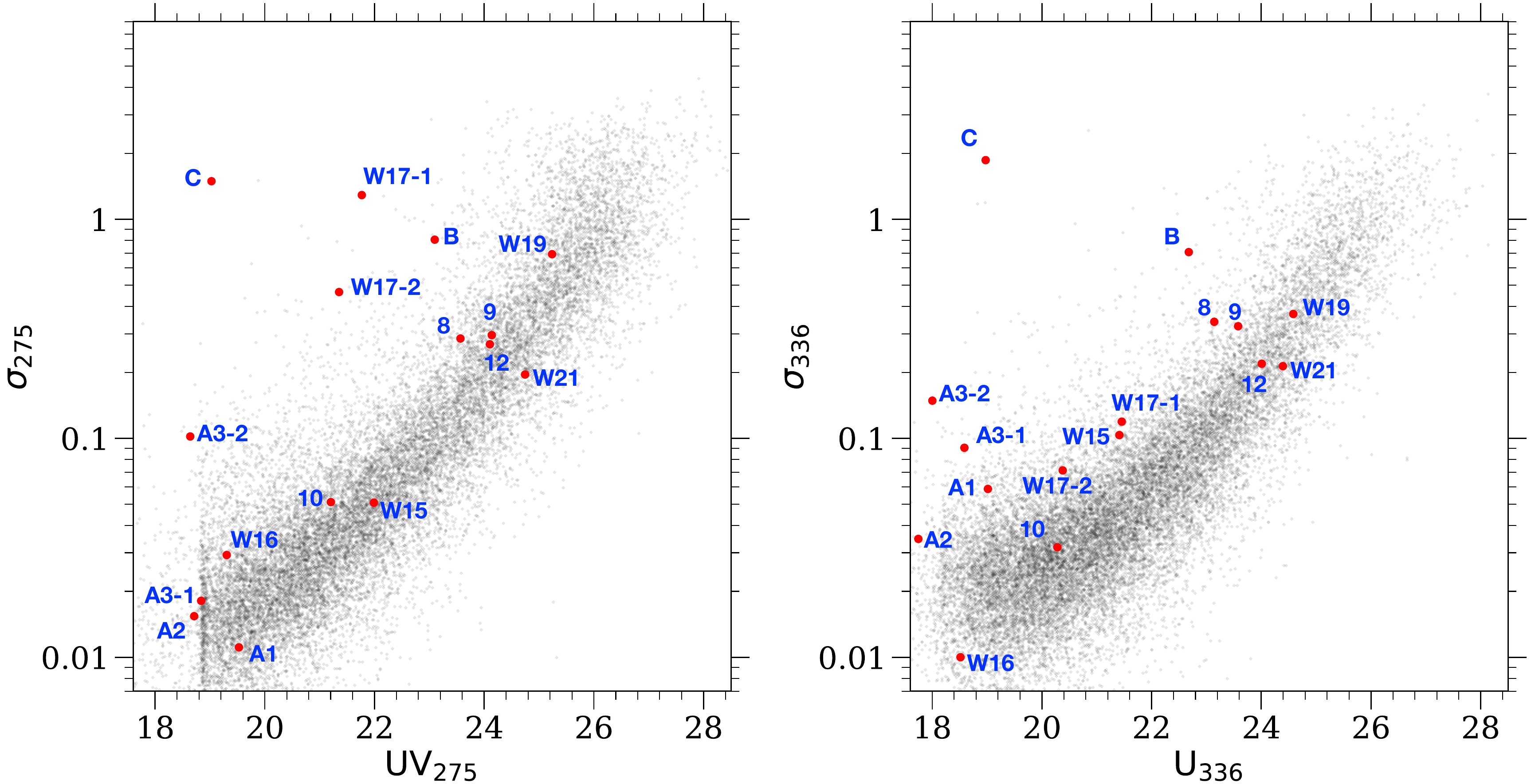}
    \caption{$\UV$ and $\U$ rms variability plotted against $\UV$ (left) and $\U$ (right) magnitudes. The locations of counterparts are indicated with filled red circles annoted with the corresponding source IDs. The majority of stars lie on a sequence, while stars with variability appear to be outliers.}
    \label{fig:uv_variability}
\end{figure*}

\begin{figure*}
    \centering
    \includegraphics[scale=0.45]{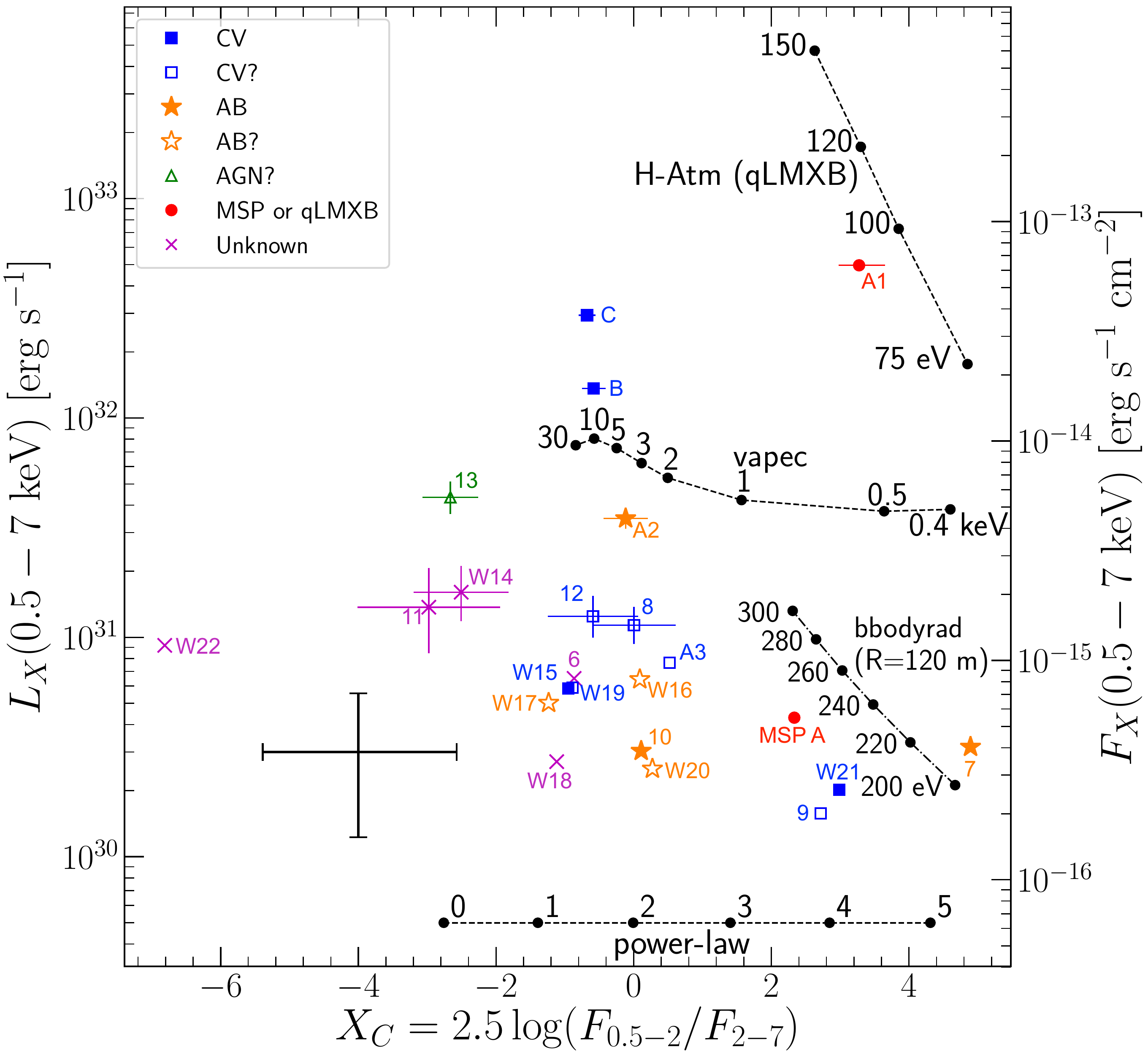}
    \caption{X-ray colour-magnitude diagram plotting X-ray hardness ratios defined in eq.(\ref{eq:x_ray_color}) vs. unabsorbed model fluxes (luminosities calculated assuming the cluster distance of $8.1~\mathrm{kpc}$) from Table \ref{tab:spectral_fits}. Different source classes are indicated with different markers, and open markers are less certain identifications. The lines indicate locations on this plot derived from models: power-law, {\tt vapec}, \cory{NS hydrogen atmosphere (H-Atm) of different temperature (using a $12~\mathrm{km}$, $1.4~M_\odot$ NS while assuming emission from the whole NS), and blackbody {\tt bbodyrad} of different temperatures assuming emission region of $120~\mathrm{m}$}. For better readability, we only plot errorbars for sources with $F(0.5-7~\mathrm{keV})\gtrsim 1.3\times 10^{-15}~\mathrm{erg~s^{-1}~cm^{-2}}$. Averaged uncertainties of the fainter sources are indicated by the capped error bars on the lower left of the plot. The fluxes of A1 are from the best-fitting atmospheric model in \citet{Echiburu20}.}
    \label{fig:x_ray_cmd}
\end{figure*}

\begin{figure}
    \centering
    \includegraphics[width=\columnwidth]{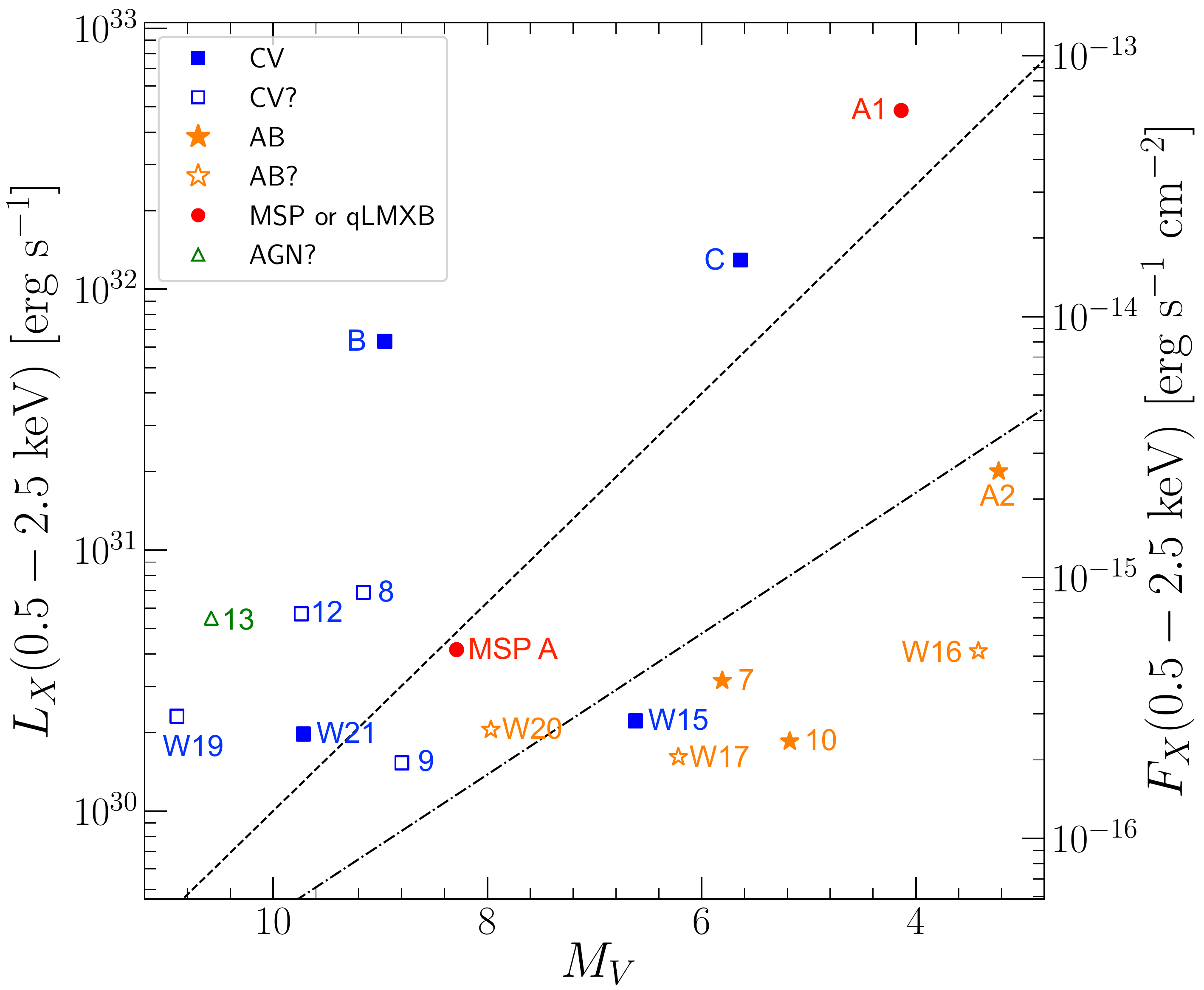}
    \caption{$0.5$--$2.5~\mathrm{keV}$ X-ray luminosities 
    vs. absolute $V$-band magnitudes ($M_V$), assuming an 8.1 kpc distance. Note that $\mathrm{V_{555}}$ magnitudes from \citet{Lugger07} are used for A2 and B to derive $M_V$s. The short-dashed line indicates the separatrix $\log_{10}L_X[0.5-2.5~\mathrm{keV}] = -0.4M_V + 34$ from \citet{Bassa04}; the dashed-dotted line marks the upper limit of $L_X$ for nearby ABs, which was derived by \citet{Verbunt08} as $\log_{10}L_X[0.5-2.5~\mathrm{keV}] = 32.3 -0.27M_V$. Open symbols are less certain identifications.}
    \label{fig:lxmv}
\end{figure}

\begin{figure}
    \centering
    \includegraphics[width=\columnwidth]{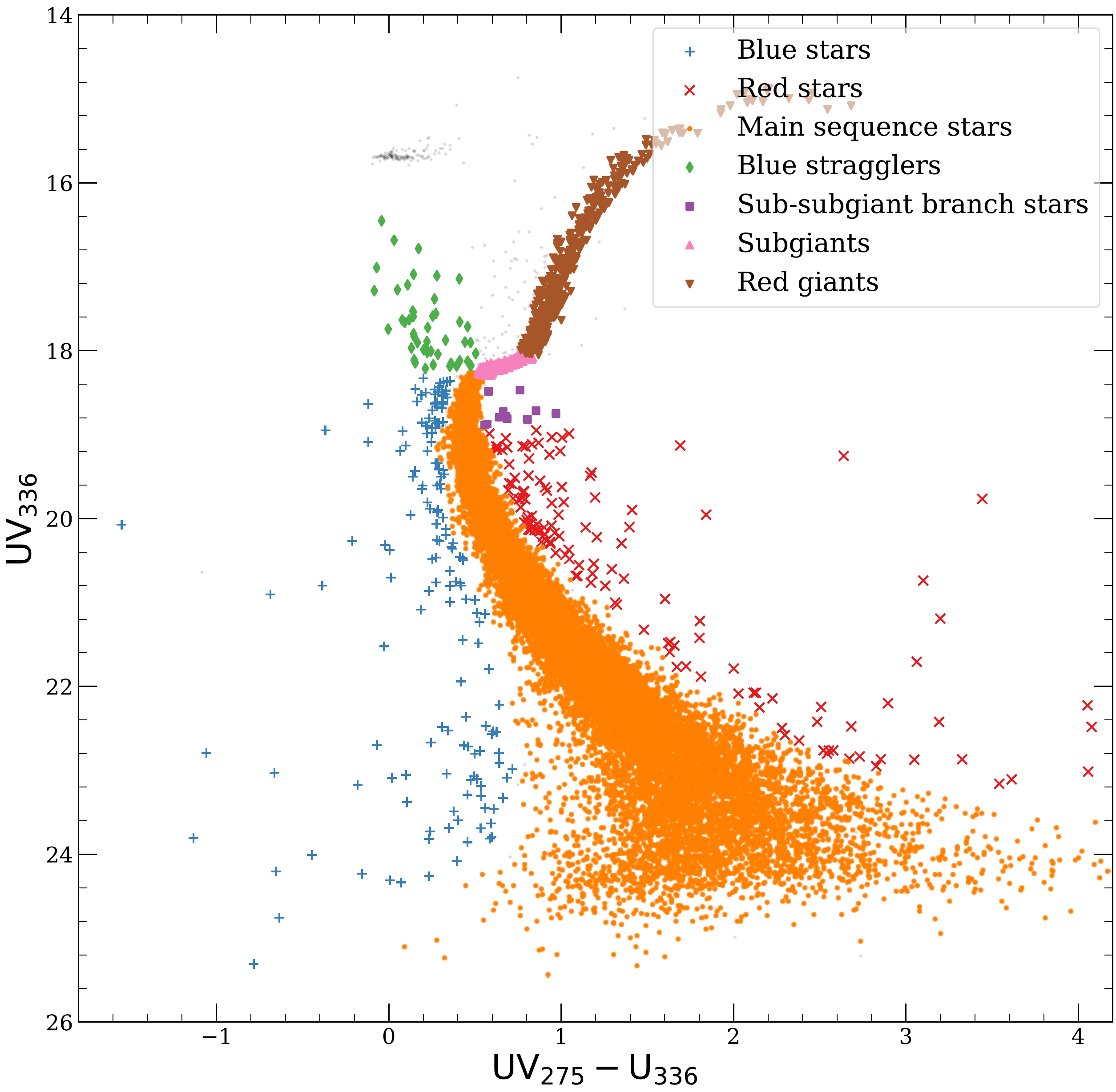}
    \caption{HUGS $\UV-\U$ CMD plotting $\U$ magnitudes vs. $\U-\UV$ colours for cluster members. Different subpopulations are plotted with different markers.}
    \label{fig:subpopulations}
\end{figure}

\begin{figure}
    \centering
    \includegraphics[scale=0.3]{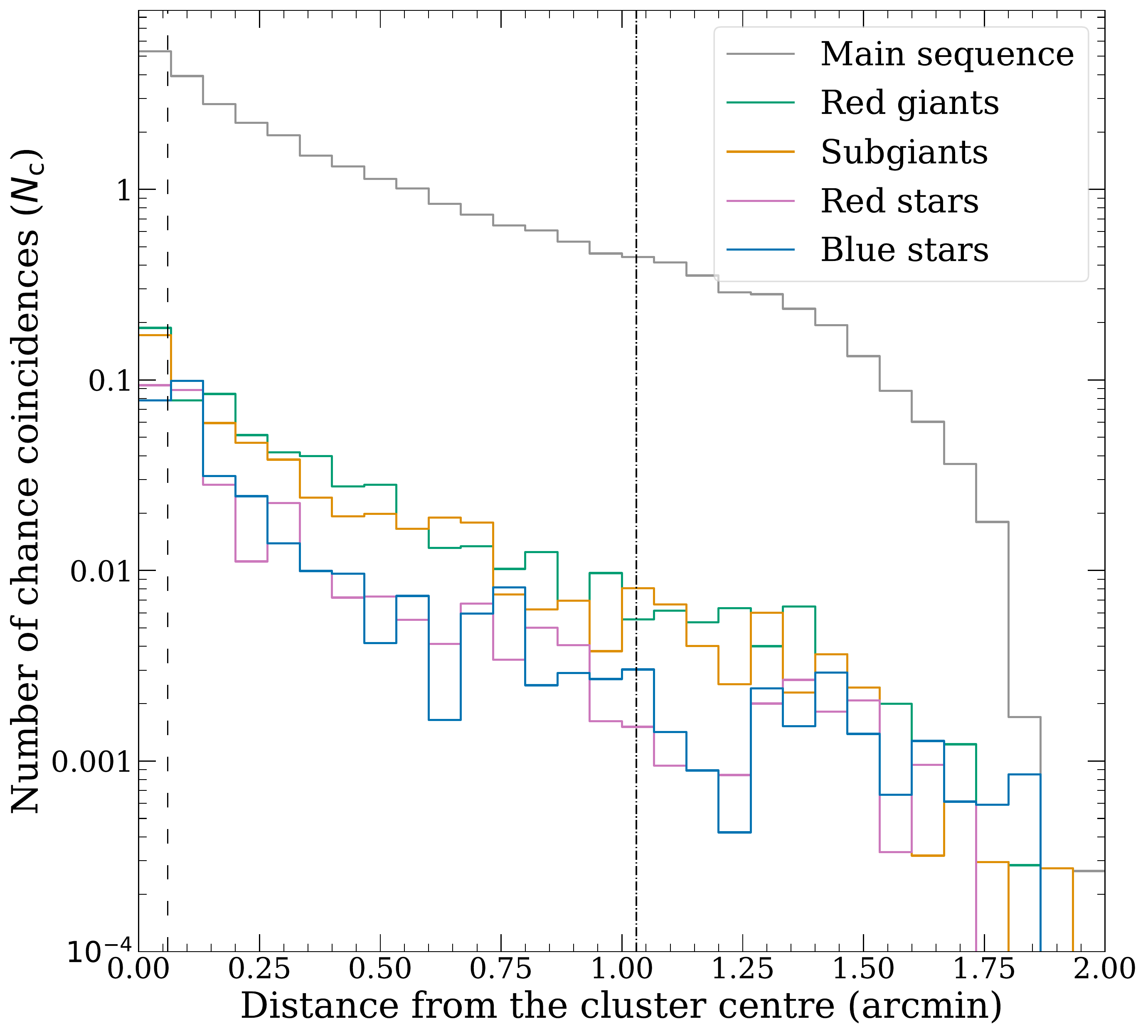}
    \caption{Number of chance coincidences ($N_\mathrm{c}$) as defined in eq. (\ref{eq:num_chance_coincidence}) vs. distance from the cluster centre (in arcmin) for different subpopulations (Section \ref{sec:chance_coincidence}). The vertical dashed and dashed-dotted lines indicate the core radius ($r_\mathrm{c}$) and half-light radius ($r_\mathrm{h}$). 
    The probability of a chance coincidence with a main sequence star is significant throughout the half-light radius.
    }
    \label{fig:chance_coincidence}
\end{figure}

\begin{table*}
    \centering
    \caption{Optical counterparts}
    \resizebox{\textwidth}{!}{
    \begin{tabular}{ccccccccccc}
    \toprule
    Source ID & HUGS ID  & $\alpha$ (ICRS) & $\delta$ (ICRS) & $\UV$ & $\U$ & $\B$ & $\V$ & $\I$ & $\mathrm{P_\mu}^a$ & Comments \\
    \midrule
    A1        & R0002625 & 21:40:22.158 & $-$23:10:46.106 & $19.53$ & $19.06$ & $19.43$ & $18.68^\dag$ & $18.14^\dag$ & $72.7\%$ &\makecell[l]{Moderate red excess \\ in $\V-\I$} \\ [0.3cm]
    A2        & -        & 21:40:22:217 & $-$23:10:47.603 & $18.70$ & $17.77$ & $17.91$ & - & - & - & \makecell[l]{Giant branch, variable \\ H$\alpha$ emission (G19)}
    \\ [0.3cm]
    A3-1      & -        & 21:40:22.044 & $-$23:10:47.348 & $18.83$ & $18.60$ & $18.77$ & - & - & - & \makecell[l]{Blue in \UV-\U,\\ red in \U-\B 
    }\\ [0.3cm]
    A3-2      & -        & 21:40:22.031 & $-$23:10:47.439 & $18.60$ & $17.93$ & $18.17$ & - & - & - & \makecell[l]{Bluer than the red \\ giant branch}\\ [0.3cm]
    B         & -        & 21:40:22.181 & $-$23:10:52.208 & $23.02$ & $22.48$ & $23.80$ & - & - & - & \makecell[l]{Moderate blue excesses \\ in UV CMDs}\\ [0.3cm]
    C         & R0039782 & 21:40:22.956 & $-$23:10:49.744 & $19.03$ & $18.97$ & $20.01$ & $20.18$ & $19.63$ & $94.3\%$ & \makecell[l]{Blue excesses \\ in the UV CMDs} \\ [0.3cm]
    6         & - & - & - & - & - & - & - & - & - & \makecell[l]{Only MS stars found \\ in the error circle} \\ [0.3cm]
    7         & R0047985 & 21:40:21.597 & $-$23:10:33.107 & $22.10$ & $21.07$ & $21.37$ & $20.35$ & $19.67$ & $96.8\%$ & \makecell[l]{Slight red excess \\ in $\V - \I$} \\ [0.3cm]
    8         & -        & 21:40:22.117 & $-$23:11:14.406 & $23.56$ & $23.15$ & $24.39$ & $23.70$ & $23.02$ & -        & \makecell[l]{Blue excesses in \\ all CMDs }\\ [0.3cm]
    9         & R0052331 & 21:40:20.434 & $-$23:10:22.722 & $24.19$ & $23.58$ & $24.36$ & $23.34$ & $21.70$ & -        & \makecell[l]{Blue excesses on \\ the UV CMDs; red  \\ excess in $\V - \I$}\\ [0.5cm]
    10        & R0061582 & 21:40:23.227 & $-$23:09:58.998 & $21.21$ & $20.28$ & $20.59$ & $19.72$ & $19.05$ & $98.1\%$ & \makecell[l]{Red excess in $\V -\I$\\ and \UV-\U}\\ [0.3cm]
    11        & - & - & - & - & - & - & - & - & - & \makecell[l]{Only MS stars found \\ in the error circle} \\ [0.3cm] 
    12        & R0044536 & 21:40:27.037 & $-$23:10:39.176 & $24.32$ & $24.15$ & $25.78$ & $24.28$ & $23.09$ & -        & \makecell[l]{ Blue in UV CMDs,\\ on MS in \V-\I} \\ [0.3cm]
    13        & R0022851 & 21:40:26.573 & $-$23:11:18.828 & -       & -       & -       & $25.12$ & $23.84$ & -        & \makecell[l]{MS counterpart consistent \\ with the {\it VLA} position} \\ [0.3cm]
    MSP A     & R0040403 & 21:40:22.404 & $-$23:10:48.957 & -       & -       & -       & $22.83^\dag$ & $21.78^\dag$ & -        & \makecell[l]{MS counterpart consistent \\ with the {\it VLA} position} \\ [0.3cm] 
    W14        & - & - & - & - & - & - & - & - & - & \makecell[l]{Only MS stars found \\ in the error circle} \\ [0.3cm] 
    W15        & R0039800 & 21:40:22.834 & $-$23:10:47.624 & $22.01$ & $21.42$ & $21.98$ & $21.16^\dag$ & $20.89^\dag$ & $96.7\%$ & \makecell[l]{Blue excesses \\ in all CMDs} \\ [0.3cm]
    W16       & R0002852 & 21:40:20.965 & $-$23:10:43.414 & $19.31$ & $18.52$ & $18.79$ & $17.96$ & $17.33$ & $97.5\%$ & \makecell[l]{A sub-subgiant} \\ [0.3cm]
    W17-1     & R0043137 & 21:40:22.177 & $-$23:10:43.350 & $22.59$ & $21.53$ & $21.69$ & $20.76^\dag$ & $20.19^\dag$ & $91.3\%$ & \makecell[l]{Slightly blue in $\V - \I$,\\ MS in UV CMDs } \\ [0.3cm]
    W17-2     & R0043073 & 21:40:22.174 & $-$23:10:43.470 & $21.03$ & $20.42$ & $20.70$ & $19.87^\dag$ & $19.30^\dag$ & $98.1\%$ & \makecell[l]{MS in all CMDs;\\ variable in UV} \\ [0.3cm]
    W18       & - & - & - & - & - & - & - & - & - & \makecell[l]{ Blended source consisting \\ of two faint MS stars} \\ [0.3cm]
    W19       & - & 21:40:23.932 & $-$23:10:13.755 & $25.14$ & $24.57$ & $26.90$ & $25.44$ & $25.00$ & - & \makecell[l]{A faint star with blue \\ excess on all CMDs; \\ large photometric errors} \\ [0.5cm]
    W20       & R0017878 & 21:40:24.143 & $-$23:11:33.304 & - & $25.89$ & $25.05$ & $22.51$ & $21.42$ & - & \makecell[l]{Moderate red excess in \\ $\V - \I$} \\ [0.3cm]
    W21       & R0046432 & 21:40:25.026 & $-$23:10:35.644 & $24.74$ & $24.40$ & $25.75$ & $24.26$ & $23.57$ & $95.3\%$ & \makecell[l]{Blue excesses in \\ all CMDs} \\ [0.3cm]
    W22       & - & - & - & - & - & - & - & - & - & \makecell[l]{Only MS stars found \\ in the error circle} \\ [0.3cm] 
    \bottomrule
    \multicolumn{11}{l}{$^a$Membership probabilities from the HUGS catalogue.}\\
    \multicolumn{11}{l}{$^\dag$ indicates magnitudes from the HUGS catalogue that are calibrated to our {\sc dolphot} photometry.}
    \end{tabular}
    }
    \label{tab:opt_counterparts}
\end{table*}

\section{Results and Discussion}
\label{sec:discussion_individual_src}
We found possible optical counterparts to 18 of the total 23 sources, of which two---MSP A and source 13---have {\it VLA} counterparts. In the following sections, we discuss these identifications in more details, incorporating information from multiple wavelengths for source classification. 

\subsection{A1---a qLMXB}
A1, the brightest cluster source, was identified by \citet{Lugger07} as a quiescent LMXB, due to the excellent fit of its X-ray spectrum to a hydrogen atmosphere NS model.
The detailed spectral fitting of A1's X-rays in \citet{Echiburu20} indicates that a normal nsatmos fit gives a remarkably small radius, apparently inconsistent with current nuclear theory.  \citet{Echiburu20} suggest two possible solutions; a) the NS  photosphere is composed of helium (also a good spectral fit to the X-ray data), in which case the companion must be a He (or hybrid) WD; b) the NS has hot spots on its surface, which would alter the inferred NS radius \citep[e.g.][]{Elshamouty16}.

\citet{Lugger07} suggested a potential counterpart to A1, which we measure to be  $0.07\arcsec$ from the nominal X-ray position. It is consistent with the main sequence on the two UV CMDs, but is redwards of the main sequence in our HST/ACS $\V-\I$ CMD (Figure \ref{fig:uv_and_vi_cmds}), although it is on the blue side of the main sequence in Lugger et al's WFPC2 $\V-\I$ CMD. 
The location of A1 in the crowded core suggests that this star may be just a chance coincidence (Figure \ref{fig:chance_coincidence}); however, the red excess in the ACS $\V-\I$ CMD may indicate an irradiated companion star, bloated by irradiation from the NS, or variability (compared to the WFPC2 $\V-\I$ CMD). Alternatively, the apparent red excess might be due to blending of a faint blue star (which is more likely to be the counterpart as suggested by \citealt{Lugger07}) and a brighter unrelated star, though this would not explain the apparent change in the two $\V-\I$ CMDs.

The X-ray information can provide some constraints on the nature of the companion star. If the X-ray spectrum is indeed produced by a helium photosphere, then the companion star must be a He (or hybrid He/C-O) WD. This would exclude the suggested main sequence optical counterpart. However, if the X-ray spectrum is distorted by the presence of hot spots, then a MS companion is possible. The similarity of this suggested companion to those of IGR J18245-2452 \citep{Pallanca13} and PSR J1740-5340 \citep{Ferraro01} raises the question of whether a transitional and/or redback MSP nature, as in those systems, is possible.  Redback and transitional MSPs show hard X-ray emission, with photon index$\sim1.0-1.5$, and $L_X\sim10^{31-32}~\mathrm{erg~s^{-1}}$ \citep[e.g.][]{Archibald10,Bogdanov10,Roberts14,Linares14,Hui14,alNoori18}, thought to be produced by an intrabinary shock between the pulsar and companion wind \citep[e.g.][]{Bogdanov05,Romani16,Wadiasingh18}.  

\citet{Echiburu20} find that the power-law component in A1's spectrum, which could be produced by an interbinary shock (or by continued accretion onto the surface), is clearly present at $L_X$(0.5-10 keV)$=1.8\times10^{31}~\mathrm{erg~s^{-1}}$ in 2017, but appears absent in 2001. We fit A1's 2001 X-ray spectrum in {\sc xspec} with a hydrogen atmosphere model plus power-law, and constrain any power-law component to $L_X<2\times10^{31}~\mathrm{erg~s^{-1}}$. A power-law component could be due to a redback MSP's shock. No redback MSP shows such a bright thermal X-ray component. An argument against a redback nature for A1 is that redback MSPs spend at least the majority of their time in a pulsar state, blowing their companion's wind out of the system, and thus should be expected to show lower time-averaged mass transfer rates than other LMXBs. However, M30 A1 has one of the highest thermal luminosities of any quiescent LMXBs \citep[e.g.][]{Heinke03}, requiring a high mass transfer rate if its thermal emission is produced by heating of the deep crust during outbursts \citep{Brown98}. Alternatively, the heating could be accomplished by more exotic mechanisms such as r-mode decay \citep{Chugunov14}, in which case M30 A1 may entirely have stopped accretion and could be active as a radio pulsar.

\subsection{Known CVs: B and C}
Both B and C were previously identified as bright CVs by \citet{Lugger07}. We confirmed the UV excesses in both $\U-\UV$ and $\U-\B$ (top panels of Figure \ref{fig:uv_and_vi_cmds}) and found that both counterparts present significant UV variability (Figure \ref{fig:uv_variability}). 
Spectroscopic MUSE observations \citep[G19 hereafter]{Gottgens19} 
identified 
broad H$\alpha$ and H$\beta$ emission features from C, indicative of an accretion disc.

\subsection{New faint CVs and candidates}
\subsubsection{A3}
A bright potential UV counterpart, A3-1, lies $0.37\arcsec$ ($\approx 0.87P_\mathrm{err}$) off the X-ray position (Figure \ref{fig:finder_p1}). The star simultaneously exhibits marked blue and red excess in the $\UV-\U$ and $\U-\B$ CMDs, respectively (top panels of Figure \ref{fig:uv_and_vi_cmds}). We note that the unusual colour combination might have been affected by light from the nearby bright star south to this star (Figure \ref{fig:finder_p1}). This bright source has a $\U$ magnitude of $15.58$ and is so saturated in the $\V$ and $\I$ images that photometry in these filters is not available for fainter stars nearby. However, this bright star does not show any sign of variability (with $\mathrm{RMS(\UV)} = 0.01$ at $\UV = 17.13$, while $\mathrm{RMS(\U)} = 0.02$ at $\U = 15.58$); the variability observed in the two fainter counterparts might not be affected by this bright object, but the WFC3 magnitudes should be interpreted with caution. Adopting this counterpart could still point to a 
CV scenario, where the 
UV excess is 
from an accretion disc and/or the WD surface, while the red excess could be partly ascribed to the companion star and/or to the moderate variability in $\U$. 

Another possible counterpart, designated A3-2, lies $0.19\arcsec$ ($\approx 0.45P_\mathrm{err}$) northeast of the X-ray position. This star resides at an uncommon location on the $\UV-\U$ CMD such that it is above the subgiant branch or slightly bluer than the red giant branch (see top panels of Figure \ref{fig:uv_and_vi_cmds}). Again, the anomalous colour could be a result of the bright nearby star, or partly accounted for by the variability we found in both $\UV$ and $\U$ (Figure \ref{fig:uv_variability}).

\subsubsection{Sources 8, 9, 12, W15, W19, and W21}
The X-ray error circle for Source 8 encloses a faint counterpart $\approx 0.11\arcsec$ ($\approx 0.31P_\mathrm{err}$) from the nominal X-ray position, which exhibits marked blue excesses in both the UV and $\V-\I$ CMDs. The UV variability analyses revealed only moderate $\U$ variability. If we adopt this counterpart, the X-ray/optical ratio will be higher than the upper limit defined for ABs (Figure \ref{fig:lxmv}), suggesting a CV nature. Although no membership information is available, we consider source 8 likely to be a cluster member based on its position not far from the cluster centre ($\approx 7.5 r_\mathrm{c}$).

\cory{We found a faint star (R0052331) in the error circle of source 9 that resides $\approx 0.23\arcsec$ ($\approx 0.54P_\mathrm{err}$) from the X-ray position. {\sc dolphot} suggests that this star exhibits moderate blue excesses in the two UVIS CMDs (top panels of Figure \ref{fig:uv_and_vi_cmds}), while showing a strong red excess in the $\V-\I$ CMD. However, the presence of the star in the $\UV$ image is not clear by visual inspection (Figure \ref{fig:finder_p2}), which leads us to be suspicious of the validity of the corresponding magnitude. As mentioned in Section \ref{sec:optical_photometry}, {\sc dolphot} uses the source list generated from the drizzle-combined $\U$ image, where R0052331 is visible, so the $\UV$ magnitude might just reflect the high background (as a result of a bright horizontal branch star northeast to R0052331) at the nominal position. The blue excess on the $\U - \B$ CMD is more robust, which leads us to our classification of source 9 as a CV candidate.}


The very faint counterpart to source 12 has photometric properties suggestive of a CV nature--it shows blue excesses in the UV (top panels of Figure \ref{fig:uv_and_vi_cmds}) while being consistent with the main sequence in the optical (bottom panel of Figure \ref{fig:uv_and_vi_cmds}). However, source 12 could be a background source, as it is 1.14' from the cluster centre, and lacks any cluster membership information.  We tentatively classify source 12 as a likely CV. 

We found a faint ($\V=21.16$) blue star in the error circle of W15, which shows blue excesses in all three CMDs (Figure \ref{fig:uv_and_vi_cmds}). This counterpart has a well-determined cluster membership ($P_\mu=96.7\%$), so we consider it a new confident cluster CV. 

Similarly, W19 
is 
consistent with a blue and UV bright object, which could indicate a CV nature. However, the membership is not determined through $P_\mu$, so W19 could also be a background AGN. We therefore classify W19 a likely CV.

Finally, the counterpart to W21 shows blue excesses on all three CMDs (Figure \ref{fig:uv_and_vi_cmds}) and has a well-determined cluster membership ($P_\mu = 95.3\%$). We thus classify it as a confident cluster CV.

\subsection{Previously suggested ABs: sources 7 and 10}
The counterpart to source 7 lies on the red side of the $\UV-\U$ main sequence and exhibits a red excess on the $\V-\I$ CMD (Figure \ref{fig:uv_and_vi_cmds}), consistent with the binary sequence. The counterpart has a well-determined cluster membership ($P_\mu = 96.8\%$). Similarly, source 10's counterpart is above the main sequence in both $\UV -\U$ and $\V - \I$ CMDs, also a cluster member ($P_\mu = 98.1\%$). Both sources were classified as cluster AB candidates by \citet{Lugger07}; now with confirmed association with the cluster, we classify them as more confident ABs.

\subsection{New ABs and candidates}
\subsubsection{A2}
The counterpart suggested by \citet{Lugger07} is a bright star that exhibits definite blue excesses on both $U-V$ and $V-I$ CMDs, 
$0.25\arcsec$ ($\approx 0.78P_\mathrm{err}$) southeast of the nominal X-ray position (Figure \ref{fig:finder_p1}). However, the VLT/MUSE spectroscopic study 
by G19 suggest another star that shows variable H$\alpha$ emission feature as the counterpart. This star lies $3$ times closer ($\approx 0.09\arcsec$ or $\approx 0.27P_\mathrm{err}$) to the X-ray position (Figure \ref{fig:finder_p1}), and is photometrically consistent with a slightly evolved red giant in the $\UV-\U$ CMD (top panels of Figure \ref{fig:uv_and_vi_cmds}). If we adopt this as the counterpart, the locus of A2 on the $L_X$-$M_V$ plane is below the separatrix (Figure \ref{fig:lxmv}), viz. X-ray fainter but optically brighter than cluster CVs, indicating an RS CVn nature. The position of A2 in the core would give rise to a moderate average number of chance coincidences with red giants ($\approx 0.2$; see Figure \ref{fig:subpopulations}); however, red giants with a variable H$\alpha$ emission feature are rare. We thus keep this star as the counterpart and conclude that A2 is an RS CVn type of AB.

\subsubsection{W16}
The counterpart to W16 is consistent with a cluster ($P_\mu =97.5\%$) sub-subgiant in $\UV-\U$ and $\V-\I$. Sub-subgiants are rare, so this counterpart is very unlikely to be a chance coincidence (Section \ref{sec:counterpart_search}). Moreover, G19 noted a variable H$\alpha$ absorption feature from this sub-subgiant, which further corroborates it as the actual counterpart. 

Sub-subgiants are typically X-ray sources with $0.5$--$2.5~\mathrm{keV}$ X-ray luminosity of $\sim 10^{30-31}~\mathrm{erg~s^{-1}}$ \citep{Geller17a}. Corresponding to the high stellar densities in the core, GC sub-subgiants are thought to be products of binary evolution or close encounters \citep{Albrow01, Geller17b}. Indeed, they have been seen in multiple source classes. For example, the MSP PSR J1740$-$5340 in the GC NGC 6397 has a well-determined sub-subgiant counterpart that exhibits clear H$\alpha$ variability \citep{Sabbi03}. \citet{Shiskovsky18} found a sub-subgiant counterpart (on their $\UV-\U$ CMD) with double-peaked H-$\alpha$ emission lines to an X-ray source in M10, also associated with a flat-to-inverted radio source. This object was considered to be a candidate accreting BH, or other exotic binary source. For W16, a BH interpretation is not likely with the absence of radio counterpart; absorption instead of emission features argue against the presence of a disc, and the lack of UV excess (top panels of Figure \ref{fig:uv_and_vi_cmds}) contradicts a CV nature. Among the common source classes in GCs, W16 is more likely an RS CVn type of AB. 


\subsubsection{W17}
\label{sec:w17}
The two possible counterparts to W17 (W17-1 and W17-2) were selected based on their marked $\UV$ variability (Figure \ref{fig:uv_variability}). These two sources are located close to each other, with W17-2 brighter in all bands (Table \ref{tab:opt_counterparts}) and closer to the X-ray position ($0.12\arcsec$ vs. $0.19\arcsec$). W17-1 appears to be consistent with the main sequence in the UVIS CMDs but shows a moderate blue excess in the $\V-\I$ CMD (Figure \ref{fig:uv_and_vi_cmds}), while W17-2 shows a slight red excess in the $\U-\B$ and $\V-\I$ CMDs. 
In the individual $\UV$ images, 
most of W17-1's PSF overlaps the chip gap in the third exposure, so its apparent variability may be spurious.


Given the uncertainty in the UV variability of W17-1 and W17-2, we have investigated the optical variability of these objects, using the WFPC2 $\Vwfpc$ images from the GO-7379 dataset. There are a total of 48 images, of which 33 are 23\,s exposures, 4 are 100\,s exposures, and 11 are 3\,s exposures. The two stars can be seen in the 23\,s and 100\,s exposures, but are not visible in the 3\,s exposures. We used {\tt FIND} and {\tt  PHOT} in {\sc daophot ii} to do aperture photometry on the 23\,s and 100\,s exposures. While {\tt FIND} is able to detect W17-2 in the 23\,s exposures, it does not detect W17-1. Thus, we added W17-1 into the object lists based on its spatial offset from W17-2. We used {\tt DAOMATCH} and {\tt DAOMASTER} to correlate the photometry from the 33 23\,s exposures and to compute the mean magnitude and $\sigma$ for each object. Similar to Figure \ref{fig:uv_variability}, in Figure \ref{fig:W17_wfpc_variability}, we present a plot of $\sigma$ vs.\ mean $\mathrm{V_{555}}$ magnitude, which gives a measure of variability. Both W17-1 and W17-2 show a deviation from the mean $\sigma$-magnitude relation, indicating variability, with the signal stronger for W17-2. The mild blue excess in the $\V - \I$ CMD might therefore be a result of this variability.

The lack of blue excess in the UV CMDs and the main sequence counterpart together hint at an AB nature for W17. We argue that W17 is likely a BY Dra system with variability caused by orbital motion.

\begin{figure}
    \centering
    \includegraphics[width=\columnwidth]{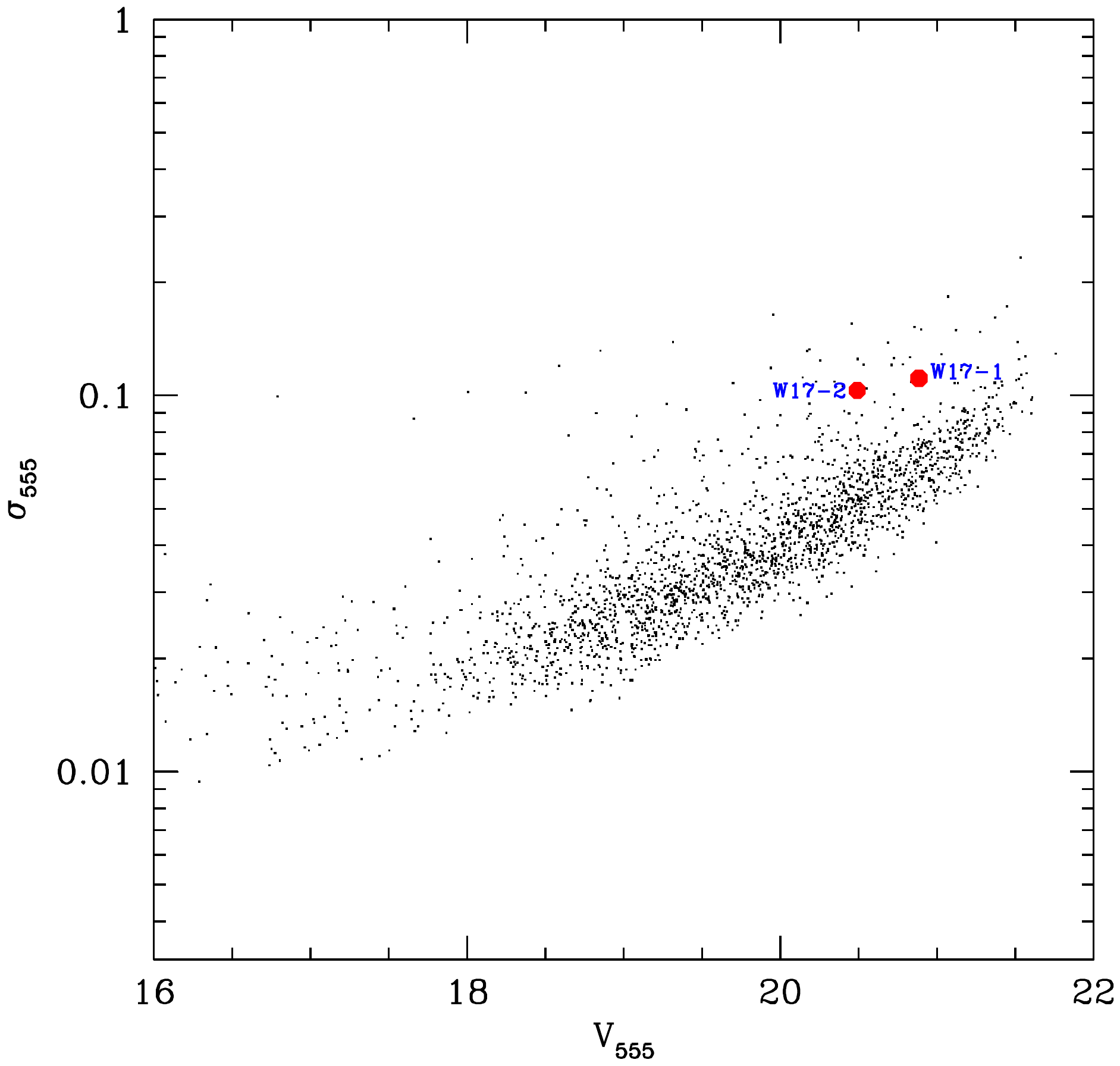}
    \caption{WFPC2 $\Vwfpc$ rms plotted versus $\Vwfpc$ magnitudes. W17-1 and W17-2 are apparently outliers to the bulk of stars, indicating marked variability.}
    \label{fig:W17_wfpc_variability}
\end{figure}



\subsubsection{W20}
W20 lies above the main sequence on the $\V-\I$ CMD, but does not have a confirmed membership, so it is also possible to be a faint foreground star. We thus classify W20 as an AB candidate.

\subsection{MSP A}
MSP A (PSR J2140$-$2310A) is an eclipsing pulsar with a spin period of $11~\mathrm{ms}$, a 4-hour orbit, and a 0.1--0.2 \msun\ companion, discovered at 
1.4$~\mathrm{GHz}$
by the {\it Green Bank Telescope; GBT} (\citealt{Ransom04}; R04 hereafter). Our $4.9~\mathrm{GHz}$ {\it VLA} observation detected a point source (ID: VLA 44) which is $0.12\arcsec$ south of the {\it GBT} timing position, with displacements ({\it VLA} $-$ {\it GBT}) in RA and DEC of $-0.016 \pm 0.028\arcsec$ and $0.116 \pm 0.051\arcsec$, respectively (uncertainties are calculated by adding the errors from the {\it VLA} and {\it GBT} in quadrature). The two observations are separated by roughly $13.7~\mathrm{yr}$, so the inconsistency of source positions might be a result of proper motion. Indeed, the displacements correspond to velocities, in RA and DEC, of $-1.18 \pm 2.03~\mathrm{mas~yr^{-1}}$ and $-8.44 \pm 3.71~\mathrm{mas~yr^{-1}}$, respectively. These are consistent with the GAIA-measured PM of the cluster ($\mu_\alpha = -0.7017 \pm 0.0063~\mathrm{mas~yr^{-1}}$, and $\mu_\delta = -7.2218\pm 0.0055~\mathrm{mas~yr^{-1}}$; see \citealt{gaia2018b}).

The $4.9~\mathrm{GHz}$ $S_\nu = 11.3~\mathrm{\mu Jy}$, but the source was not detected in the $7~\mathrm{GHz}$ band ($S_\nu < 4.8~\mathrm{\mu Jy}$, at the $3~\sigma$ level), so we only get a very rough constraint on the spectral index: $\alpha<0$.

We also detect 25 counts from an X-ray source $0.23\arcsec$ from the {\it VLA}  position (previously suggested by R04 on the basis of 5 counts). 
The X-ray spectrum can be modelled with an absorbed {\tt bbodyrad} or  {\tt pow} model. The former yields a temperature of $0.3\pm0.1~\mathrm{keV}$ but an unconstrained emission radius; the {\tt pow} model results in a photon index $\Gamma = 2.9 \pm 0.9$,  consistent with the photon index range observed in faint MSPs
(Table \ref{tab:spectral_fits}).
Indeed, either model should be physically possible as we expect both thermal emission from the NS surface, and non-thermal emission resulting from accelerated particles between the interacting pulsar and companion winds \citep[see e.g.,][]{Harding90, Arons93, Romani16}. The inferred blackbody parameters, and 0.5-10 keV $L_X$ of $4\times10^{30}$ erg s$^{-1}$, are consistent with thermal emission from MSPs in other GCs \citep{Bogdanov06,Forestell14}.

Our {\sc dolphot} ACS analysis found the suggested counterpart in both $\V$ and $\I$, yielding magnitudes of $22.83\pm 0.02$ and $21.78\pm 0.07$, respectively (the errors are calculated by adding the intrinsic {\sc dolphot} error and the calibration error, described in Section \ref{sec:optical_photometry}, in quadrature). Our photometry places the star slightly redward of the main sequence on the $\V-\I$ CMD (bottom panel of Figure \ref{fig:uv_and_vi_cmds}), confirming that the companion is not degenerate.
The slight red excess 
may be due to bloating of the companion star during mass transfer \citep{Knigge11}.


R04 inspected 1676 s WFPC2 observations (GO-7379) and reported $\mathrm{V_{555}}$ detections and $\I$ nondetections of a faint counterpart, indicating that the object is either on or bluer than the main sequence. We re-examined the co-added WFPC images, where the faint counterpart is only visible by visual inspection, so our {\sc daophot} routine (Section \ref{sec:w17}) did not measure its magnitudes. We therefore manually run aperture photometry on the faint star, finding $\mathrm{V_{555}} - \mathrm{I_{814}} = 0.8\pm 0.6$. This is consistent with the ACS result.

\subsection{Unclassified sources and likely AGN}
The sources that do not have a likely counterpart are 6,  11, W14, W18, and W22. These sources only have stars that are consistent with the main sequence in one or multiple CMDs. We leave them unclassified in our catalogue but tentatively discuss their possible nature in this section.

The two main sequence stars in source 6's error circle may be chance coincidences, considering the relatively high number of chance coincidences ($\approx 2$) expected near the cluster centre (Figure \ref{fig:chance_coincidence}). Fitting an {\tt apec} model to source 6's X-ray spectrum does not constrain $N_\mathrm{H}$, so we fixed it to the cluster value (Section \ref{sec:data_analyses}). 
We note a likely emission feature at $\sim 1.9~\mathrm{keV}$ (Figure \ref{fig:spec_6}), which does not match any emission feature at the cluster abundances. One possible explanation is that source 6 is extragalactic, so this line could be a redshifted Fe K-$\alpha$ emission line. 

Compared to source 6, W18 and W22 are farther from the cluster core, so their main sequence counterparts are less likely to be chance coincidences ($N_\mathrm{c}\approx 0.7$ and $\approx 1$ for W18 and W22, respectively). 
There are three sources in W18's error circle, two blended together, all of which are consistent with the main sequence. If the two stars were associated, we might further classify W18 a likely BY Dra type of AB.

The 3 stars in W22's error circle are also consistent with the main sequence. However, unlike W18, W22 has a very hard X-ray spectrum (no counts in $0.5-2~\mathrm{keV}$; Figure \ref{fig:x_ray_cmd}) 
which is not seen in 
faint cluster ABs. We thus suspect that none of the main sequence stars are the true counterpart, and consider W22 a likely AGN. \cory{Sources 11 and W14 are also rather spectrally hard (Figure \ref{fig:x_ray_cmd}), suggesting enhanced absorption and thus likely an extragalactic nature. 
}

There is a more definite AGN candidate---source 13---which has a steep-spectrum ($\alpha = -1.44^{+0.16}_{-0.17}$) radio counterpart (VLA6) and, like sources 11, W14 and W22, shows a very hard X-ray spectrum. There is a very faint optical counterpart (within the radio error circle) that is marginally detected in the $\V$ and $\I$ bands (Figure \ref{fig:finder_p1}). Another nearby radio source, VLA29, lies $1.6\arcsec$ ($\approx 1.7P_\mathrm{err}$) southwest of source 13's X-ray position. VLA29 may also be associated with X-ray source 13 and VLA6, such as a radio lobe from a central AGN. The optical source appears to be consistent with the main sequence (bottom panel of Figure \ref{fig:uv_and_vi_cmds}); however, the  radio-optical match with a tiny radio error circle suggests that this counterpart is very unlikely to be a chance coincidence. Overall, the hard X-ray spectrum, large offset from the cluster center, and radio counterpart together hint at an extraglactic nature of source 13.

Finally, we can estimate the predicted number of AGN in our field, and compare it with the number of likely AGN we see. 
We apply a $0.5$--$2~\mathrm{keV}$ flux limit of $1.1\times 10^{-16}~\mathrm{erg~s^{-1}~cm^{-2}}$  (from the faintest source detectable by {\tt wavdetect} on the $0.5$--$2~\mathrm{keV}$ image given the threshold parameter in Section \ref{sec:data_analyses}). The expected number of AGN within a radius of $1.15\arcmin$ (our search radius) is $4^{+4}_{-2}$ according to the empirical formula from  \citet{Mateos08}\footnote{The errors correspond to 90\% confidence levels, according to \citet{gehrels1986}.}; this number is consistent with the number of likely AGN (6, W22, 11, 13, W14, and W22) we have found in our catalogue.

\begin{figure}
    \centering
    \includegraphics[scale=0.55]{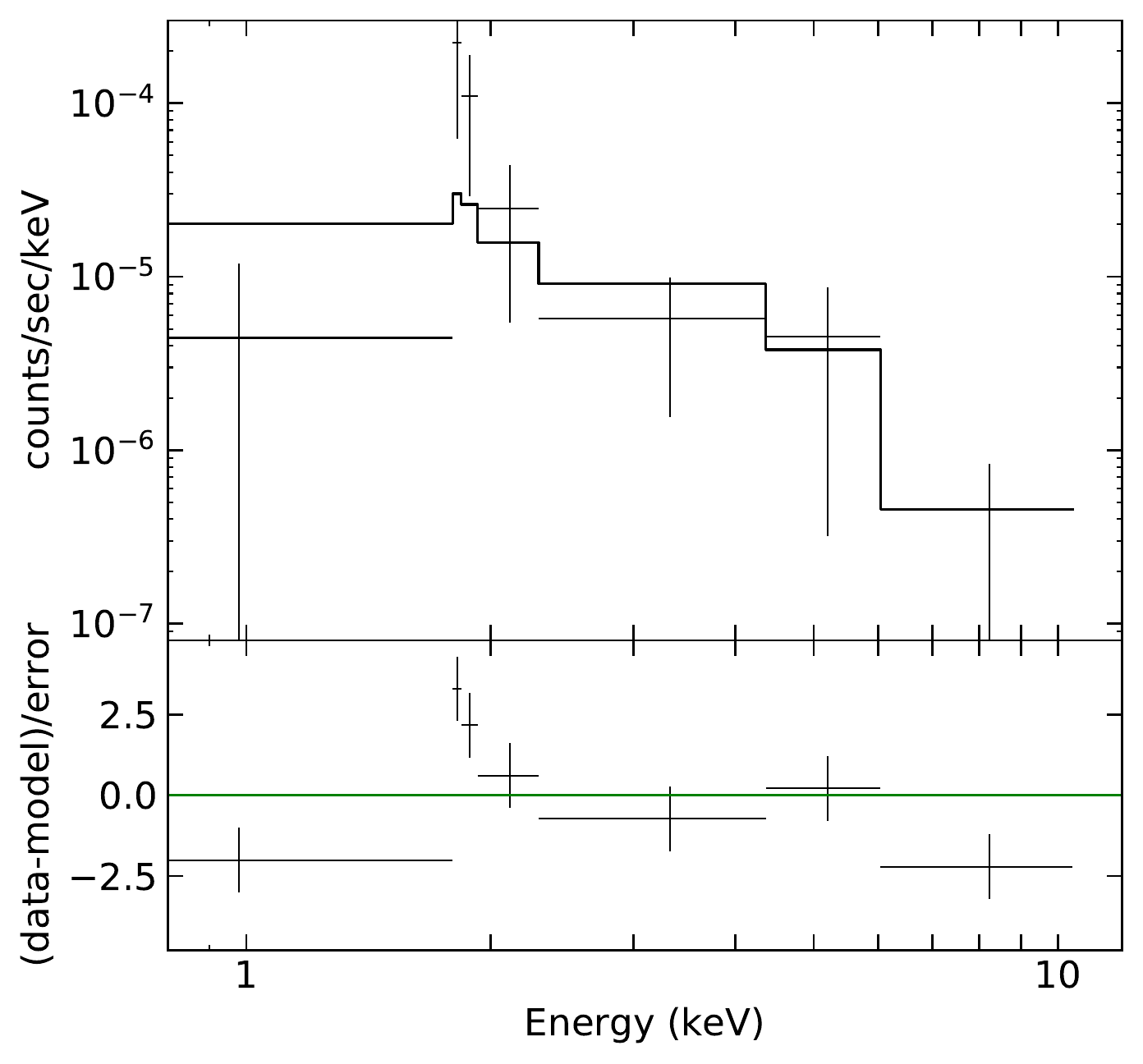}
    \caption{{\it Top}: X-ray spectrum of source 6 fitted with a {\tt vapec} model (the spectrum is rebinned for better readability). {\it Bottom}: fitting residuals; a likely emission feature is present around $1.9~\mathrm{keV}$.}
    \label{fig:spec_6}
\end{figure}

\subsection{Spatial distribution of different source classes}
\label{sec:spatial_distribution}
In order to study the spatial distribution of the \emph{Chandra} sources detected in M30, we followed the analysis methods of \cite{cohn10} and \cite{Lugger17}. We first fit a ``generalised King model'' (also known as a cored power law) of the form,
\begin{equation}
S(r) = S_0 \left[1 + \left({\frac{r}{r_0}}\right)^2 \right]^{\alpha/2},
\label{eqn:Cored_PL} 
\end{equation}
to a main sequence turnoff (MSTO) sample, defined as stars in the HUGS database within 0.5 mag in $\V$ of the MSTO magnitude. We adopted the HUGS determination of the cluster centre. We chose the HUGS star counts for this analysis given their greater apparent completeness, relative to the {\small DOLPHOT} counts, near the cluster centre. While we fit the unbinned star counts, we have visualised the binned radial profile with the model fit in Figure~\ref{f:radial_profile}. The central cusp in the profile is visible as a correlated deviation from the model fit within about 20\arcsec\ of the cluster centre. Nonetheless, the model gives a reasonable approximation to the profile within the half-light radius and provides a basis for comparison with other object groups. 

We next considered several different source samples including all CVs plus the qLMXB A1, bright CVs, faint CVs, and ABs, in order to determine the characteristic masses for these samples. In defining the bright and faint samples, we may either use the optical magnitude or the X-ray flux. We have chosen the former in our previous work on NGC~6397 and NGC~6752 \citep{cohn10,Lugger17}. Doing so resulted in two well-defined samples in each cluster. We note that the identifications in M30 of A1 (the qLMXB) and A3 with bright optical objects are less certain than the other identifications. In any case, these two objects have high X-ray fluxes, so would be placed in the bright group if the selection were alternatively based on X-ray flux. Using $B_{438}$ magnitude, the bright CV group is comprised of A1, A3-1, C, and W15, and the faint CV group is comprised of B, 8, 9, W19, and W21. We note that we excluded W12, since it lies outside of the half-light radius and thus has a higher likelihood of being an AGN. As can be seen in Figure~\ref{f:cumulative_radial_distributions}, the bright CVs have a more centrally concentrated distribution than the MSTO sample, whereas the other groups are not more centrally concentrated. This is quantified by Kolmogorov-Smirnov (K-S) comparisons of the samples to the MSTO sample which are listed in Table~\ref{t:Cored_PL_fits}. Only the bright CV group differs from the MSTO sample at a significant level, $p<1\%$. A direct K-S comparison of the bright and faint CV samples indicates that they are inconsistent at the 5.3\% level, which is marginally significant. 

To estimate the characteristic masses of the individual groups, we fit each group with a surface density profile of the form,
\begin{equation}
\label{eqn:Cored_PL_component} 
S(r) = S_0 \left[1 + \left({\frac{r}{r_0}}\right)^2 \right]^{[q(\alpha_\mathrm{to}-1)+1]/2},
\end{equation}
by maximising the likelihood over $q = m/m_\mathrm{to}$, where $m_\mathrm{to} = 0.80\,\msun$ is the assumed MSTO mass \citep{Lugger07}, and $r_0$ and $\alpha_\mathrm{to}$ are determined by the previous fit to the MSTO group. The results of the fits are given in Table~\ref{t:Cored_PL_fits}. The $q$ values for the all CV, bright CV, and AB groups exceed unity, with the significance ratio expressed in $\sigma$. The $q$ value excess above unity is significant only for the bright CV sample, at the $2\,\sigma$ level. For the faint CV group, $q$ is less than unity, although not at a significant level.

As we found for NGC~6397 \citep{cohn10} and NGC~6752 \citep{Lugger17}, the bright CVs in M30 (in this case including one qLMXB) are more centrally concentrated than the MSTO stars, while the faint CVs and ABs are not. The implied mass for the bright CVs in M30, $1.5 \pm 0.3\,\msun$, is consistent with what we found for these other two clusters. We note that the determination, in \citet{Heinke03}, of the mean mass of 20 qLMXBs in seven clusters resulted in a value of $1.5 \pm 0.2\,\msun$. Thus, the inclusion of the qLMXB A1 in the bright CV sample for M30 should not bias the determination of the typical bright CV mass. Indeed, if A1 is excluded from the bright CV sample, the mean mass is found to be $1.3 \pm 0.3\,\msun$, with a mass excess above the MSTO mass significant at the $1.9\,\sigma$ level. 

As we have discussed for NGC~6397 and NGC~6752, the finding that the bright CVs are more centrally concentrated than the faint CVs is consistent with the bright CVs representing a recently formed population that is produced by dynamical interactions near the cluster centre \citep{cohn10,Lugger17}. 
As recently formed CVs age, the mass of the secondary decreases and the accretion rate declines, leading to a reduction in both the optical and X-ray luminosity \citep{Howell01}.
The observation of a double blue straggler sequence in M30 provides independent evidence for a recent core collapse event that has resulted in the production of dynamically formed populations \citep{Ferraro09}. 
However, \citet{Belloni19} argue that their simulations of CVs in GCs push CVs out of the core in the dynamical interaction that forms them, and prefer a mass segregation argument for the higher concentration of bright CVs. 

\begin{table*}
\caption{Cored-Power-Law Model Fit Results} 
\label{t:Cored_PL_fits}
\begin{tabular}{lrcccccr}
\hline
Sample & 
$N^a$ &
$q$ & 
$r_c~(\arcsec)$ &
$\alpha$ &
$m~(\msun)$ &
$\sigma^b$ &
K-S prob$^c$ \\
\hline
%
%
MSTO       & 1881 &   1.0            & $ 4.4 \pm 0.9$ & $-1.41 \pm 0.06$ & $0.80 \pm 0.05$ & \nodata & \nodata \\
all CV     &    9 &  $1.28 \pm 0.19$ & $ 3.3 \pm 0.5$ & $-2.09 \pm 0.47$ & $1.02 \pm 0.15$ & 1.5     & 10\%    \\
bright CV  &    4 &  $1.85 \pm 0.42$ & $ 2.4 \pm 0.4$ & $-3.45 \pm 1.01$ & $1.48 \pm 0.34$ & 2.0     & 0.68\%   \\
faint CV   &    5 &  $0.90 \pm 0.27$ & $ 5.2 \pm 4.9$ & $-1.16 \pm 0.65$ & $0.72 \pm 0.22$ & $-0.4$  & 55\%    \\
AB         &    6 &  $1.20 \pm 0.24$ & $ 3.6 \pm 0.9$ & $-1.89 \pm 0.58$ & $0.96 \pm 0.19$ & 0.8     & 42\%    \\
\hline
\multicolumn{8}{l}{\makecell[tl]{$^a$Size of sample within 61\farcs8 of cluster centre}}\\
\multicolumn{8}{l}{\makecell[tl]{$^b$Significance of mass excess above MSTO mass in sigmas}}\\
\multicolumn{8}{l}{\makecell[tl]{$^c$K-S probability of consistency with MSTO group}}\\
\end{tabular}
\end{table*}

\begin{figure}
    \centering
    \includegraphics[width=\columnwidth]{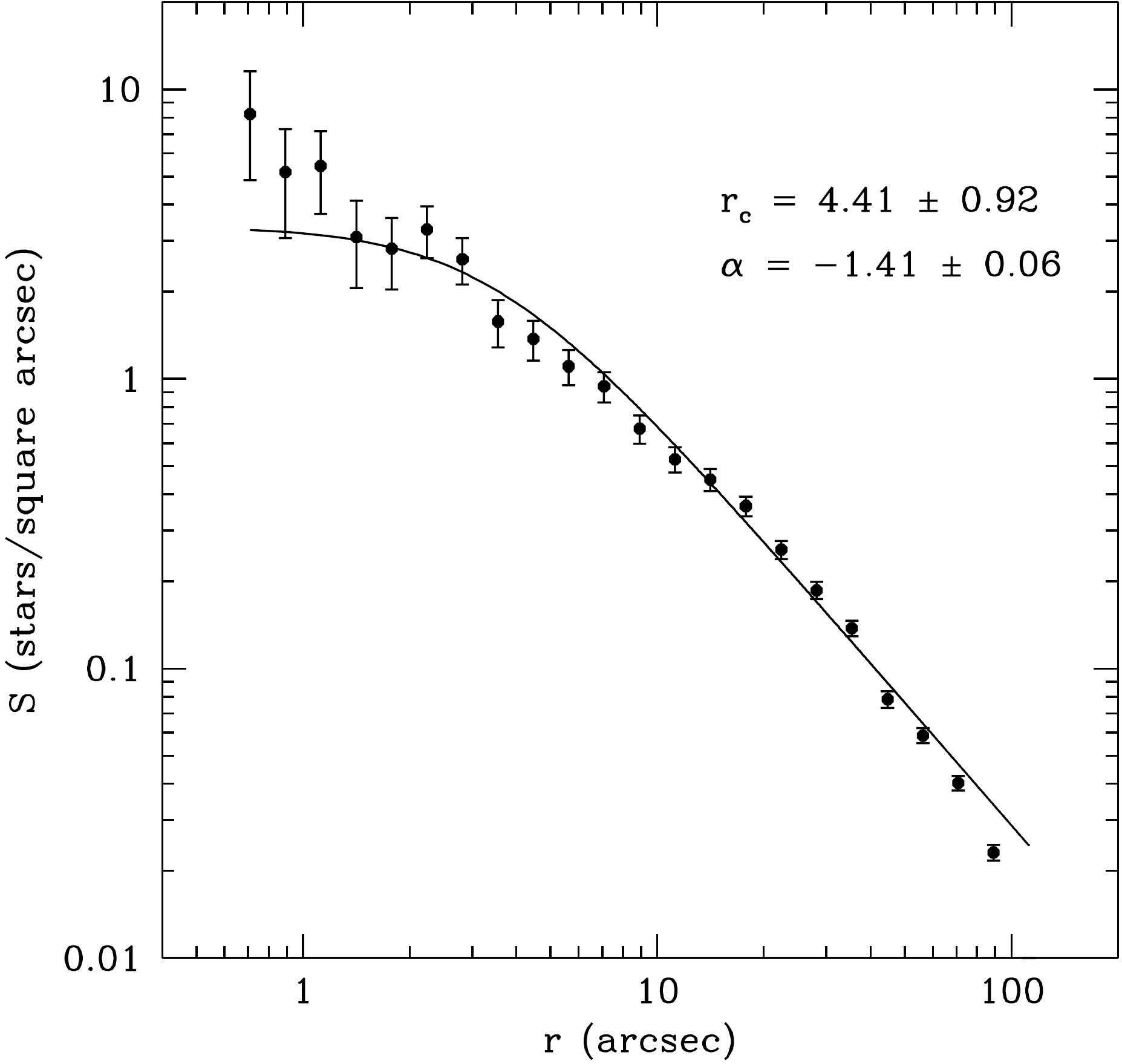}
    \caption{Binned radial profile of MSTO sample with a cored power law fit out to the half-light radius.}
    \label{f:radial_profile}
\end{figure}

\begin{figure}
    \centering
    \includegraphics[width=\columnwidth]{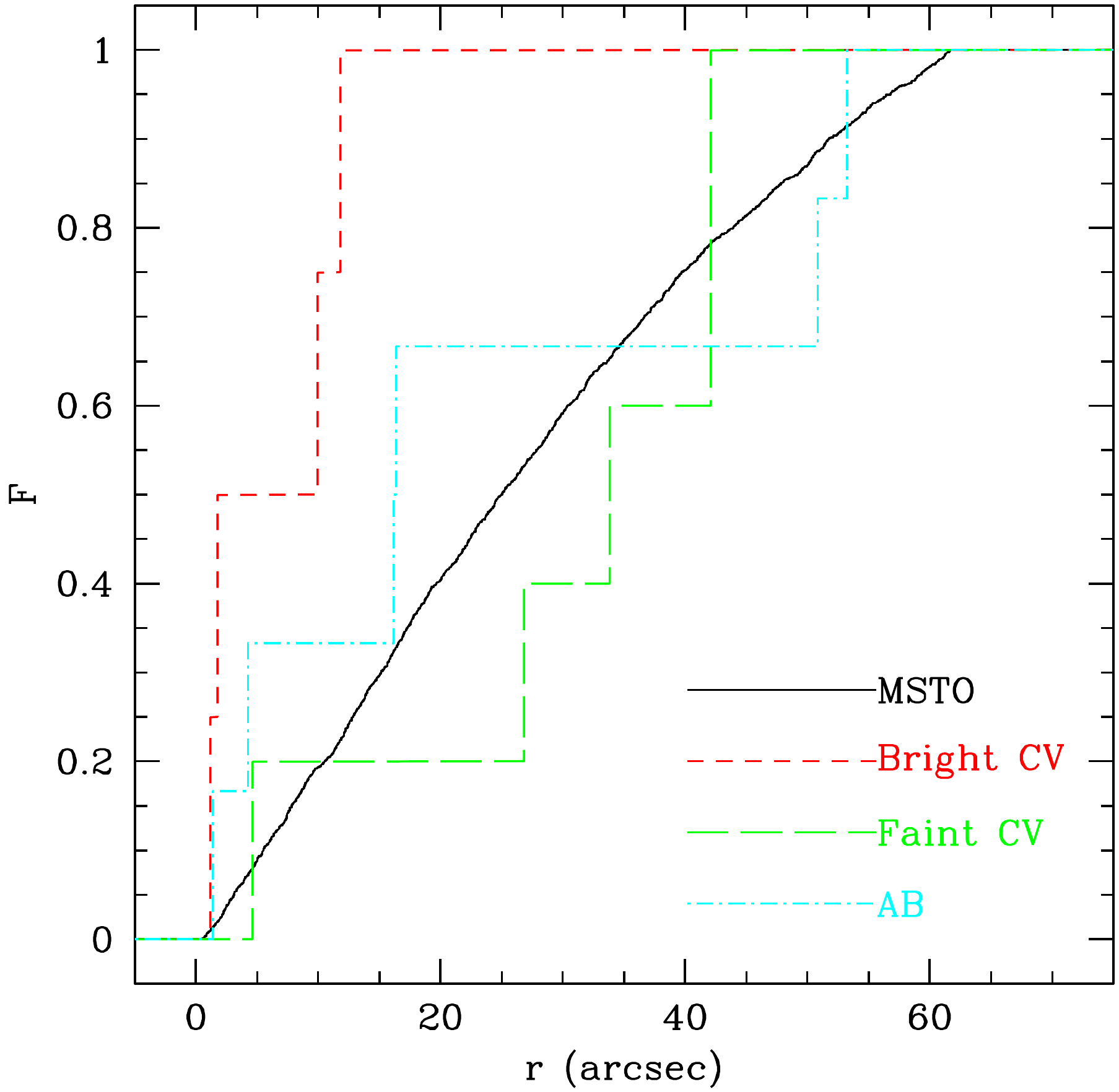}
    \caption{Cumulative radial distributions for selected stellar groups. Fitting information and K-S sample comparisons for these stellar groups are given in Table~\ref{t:Cored_PL_fits}.}
    \label{f:cumulative_radial_distributions}
\end{figure}

\section{Conclusions}
\label{sec:conclusions}
Our deep X-ray observation revealed a total of 10 new X-ray sources within the half-light radius, extending the original catalogue of \citet{Lugger07} to 23 X-ray sources. Comparing X-ray positions with UV, V, and I band {\it HST} observations, we found optical counterparts to $18$ of the $23$ sources, identifying $2$ new CVs (W15 and W21), $5$ new CV candidates (A3, 8, 9, 12, W19), two RS CVn candidates (A2 and W16), and $2$ new AB candidates (W17 and W20). Cross-matching the {\it VLA} catalogue with our {\it Chandra} catalogue revealed two matches---MSP A and 13---which also match with their optical counterparts. The counterpart to MSP A lies slightly redward of the main sequence, which we interpret as a result of stripped mass from the companion star. The radio counterpart to source 13 matches a faint optical source  consistent with the main sequence; but we classify 13 as a likely AGN based on its hard X-ray spectrum. The remaining 5 sources (6, 11, W14, W18, and W22) do not have definite optical counterparts, so we tentatively classify them to be extragalactic (6, 11, W14, and W22), or a faint cluster AB (W18). 

Based on our classification, we performed a K-S comparison of the radial profiles of bright and faint populations. We found that  bright CVs are more centrally concentrated than faint CVs. This is consistent with other core-collapsed GCs.


\section*{Acknowledgements}

COH \& GRS acknowledge NSERC Discovery Grants RGPIN-2016-04602 and RGPIN-2016-06569 respectively, and COH also a Discovery Accelerator Supplement. JS acknowledges support from a Packard Fellowship and NSF grants AST-1308124 and AST-1514763.  
The National Radio Astronomy Observatory is a facility of the National Science Foundation (NSF) operated under cooperative agreement by Associated Universities, Inc. (AUI).
This research has made use of data obtained from the Chandra Data Archive and the Chandra Source Catalog, and software provided by the Chandra X-ray Center (CXC) in the application package CIAO.
Optical analyses in this work are based on observations made with the NASA/ESA Hubble Space Telescope, obtained from the data archive at the Space Telescope Science Institute. STScI is operated by the Association of Universities for Research in Astronomy, Inc. under NASA contract NAS 5-26555.
This work has also used data from the European Space Agency (ESA) mission {\it Gaia} (\url{https://www.cosmos.esa.int/gaia}), processed by the {\it Gaia}
Data Processing and Analysis Consortium (DPAC,
\url{https://www.cosmos.esa.int/web/gaia/dpac/consortium}). Funding for the DPAC
has been provided by national institutions, in particular the institutions
participating in the {\it Gaia} Multilateral Agreement.

\section*{Data Availability}
The {\it Chandra} data used in this article are available in the Chandra Data Archive (\url{https://cxc.harvard.edu/cda/}) by searching the Obs. ID listed in Table \ref{tab:x_ray_obs} in the ChaSeR interface (\url{https://cda.harvard.edu/chaser/}). The {\it HST} data can be retrieved from the Mikulski Archive for Space Telescopes (MAST) Portal (\url{https://mast.stsci.edu/portal/Mashup/Clients/Mast/Portal.html}) with the proposal IDs listed in Table \ref{tab:hst_obs}. The HUGS catalogue used in this work can be downloaded from \url{https://archive.stsci.edu/prepds/hugs/}. VLA observations can be downloaded from the VLA data archive (\url{https://science.nrao.edu/facilities/vla/archive/index}) by searching the program ID: 15A-100.




\bibliographystyle{mnras}
\bibliography{ref} 




\appendix
\section{All finding charts}
\begin{figure*}
    \centering
    \includegraphics[width=\textwidth]{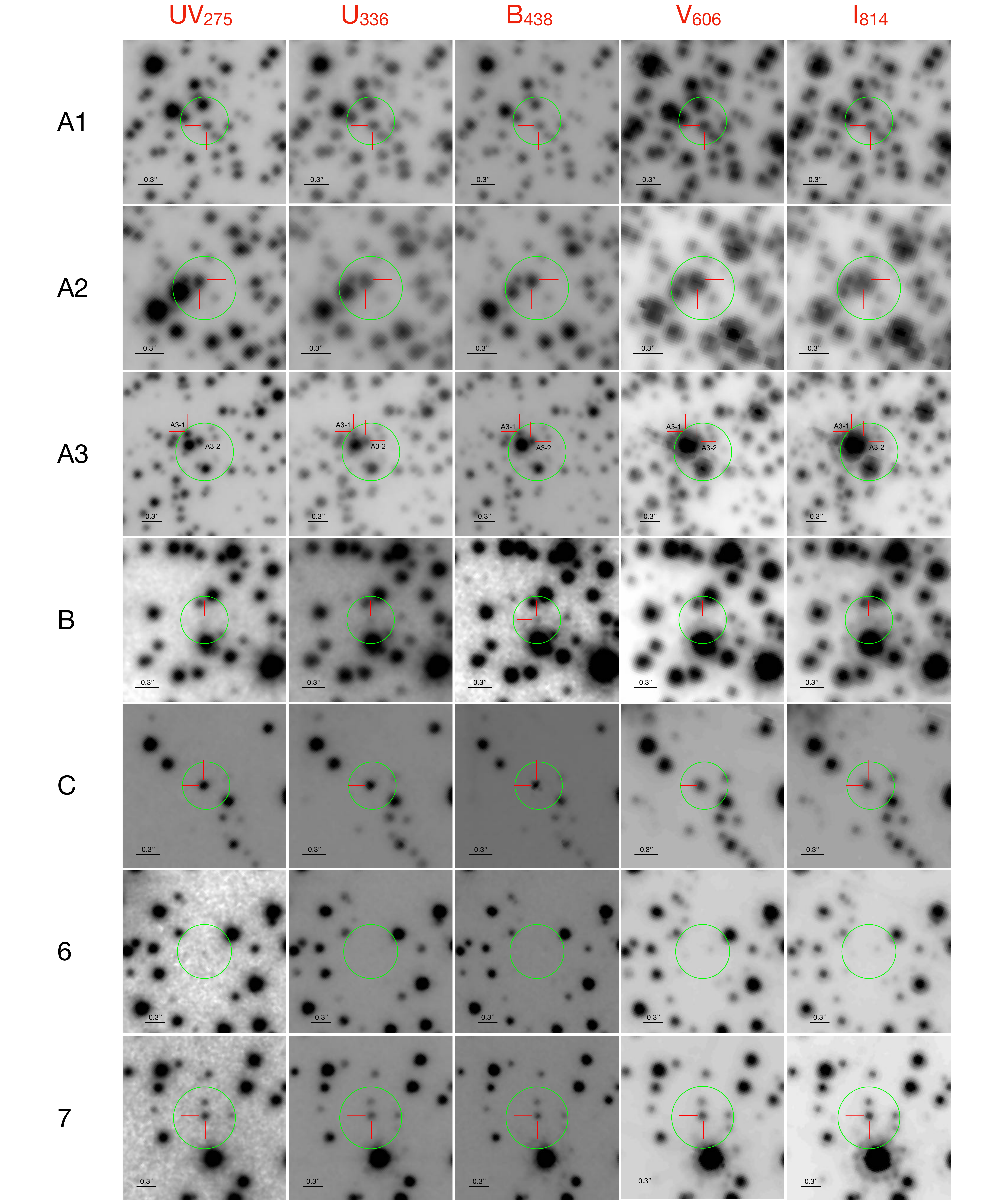}
    \caption{$\UV$, $\U$, $\B$, $\V$, and $\I$ finding charts for $23$ X-ray sources; north is up and east is to the left. The green circle in each finding chart marks the $95\%$ {\it Chandra} error circle as per \citet{hong2005}, and identified counterparts are indicated with red cross hairs. In the charts for source 13 and MSP A, the ellipses in cyan marks the $1~\sigma$ radio positional uncertainty.}
    \label{fig:finder_p1}
\end{figure*}

\begin{figure*}
    \centering
    \includegraphics[width=\textwidth]{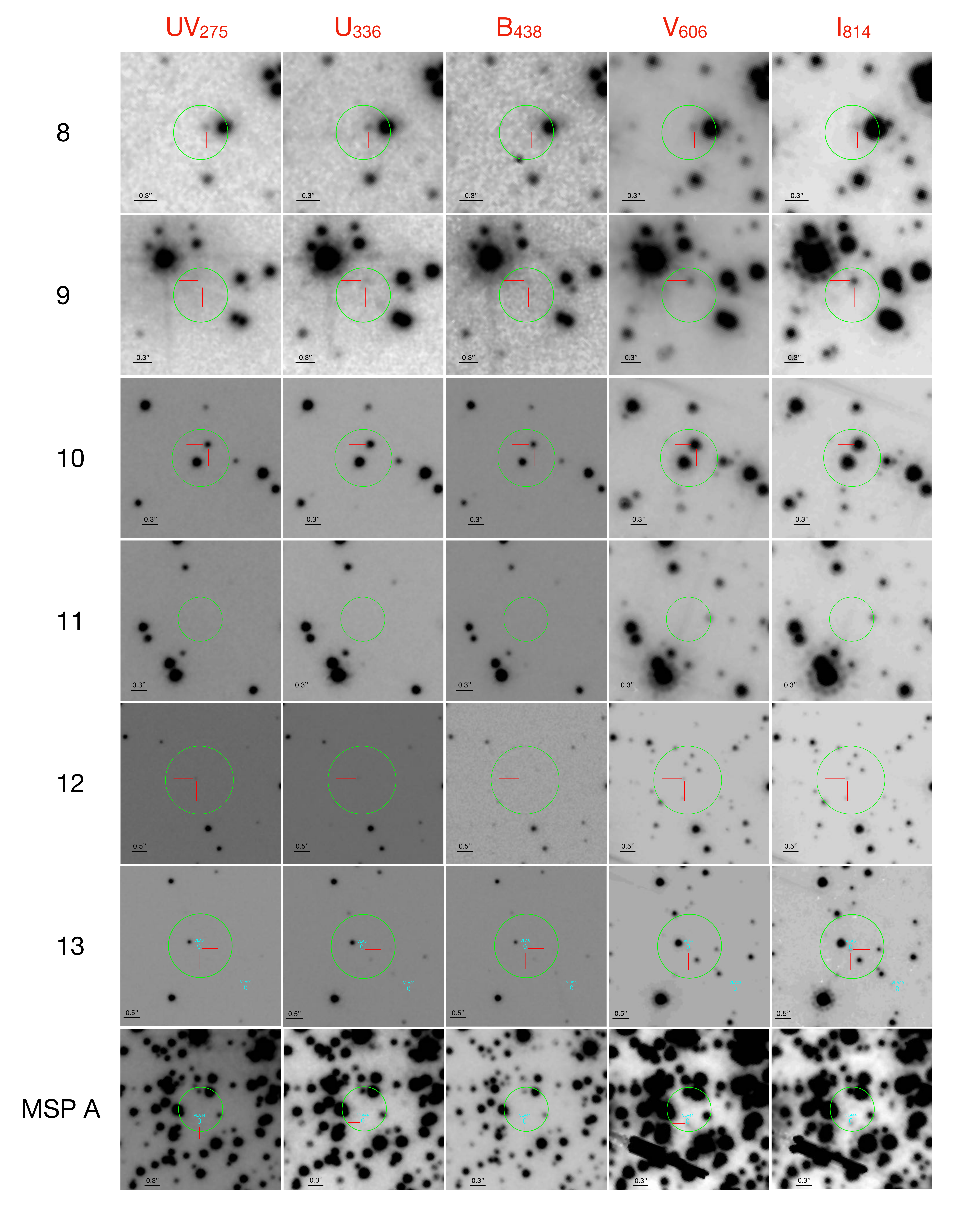}
    \contcaption{}
    \label{fig:finder_p2}
\end{figure*}

\begin{figure*}
    \centering
    \includegraphics[width=\textwidth]{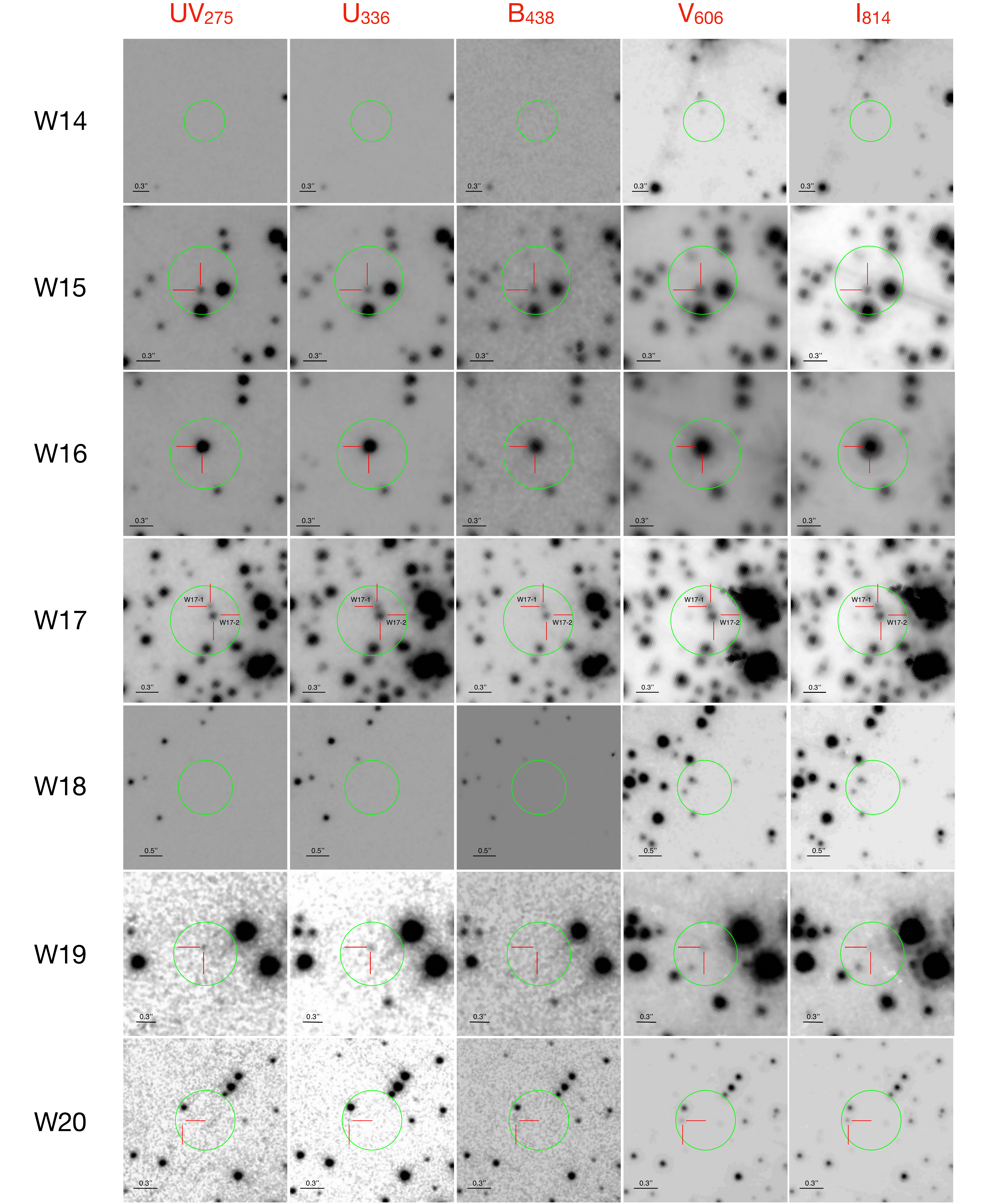}
    \contcaption{}
    \label{fig:finder_p3}
\end{figure*}

\begin{figure*}
    \centering
    \includegraphics[width=\textwidth]{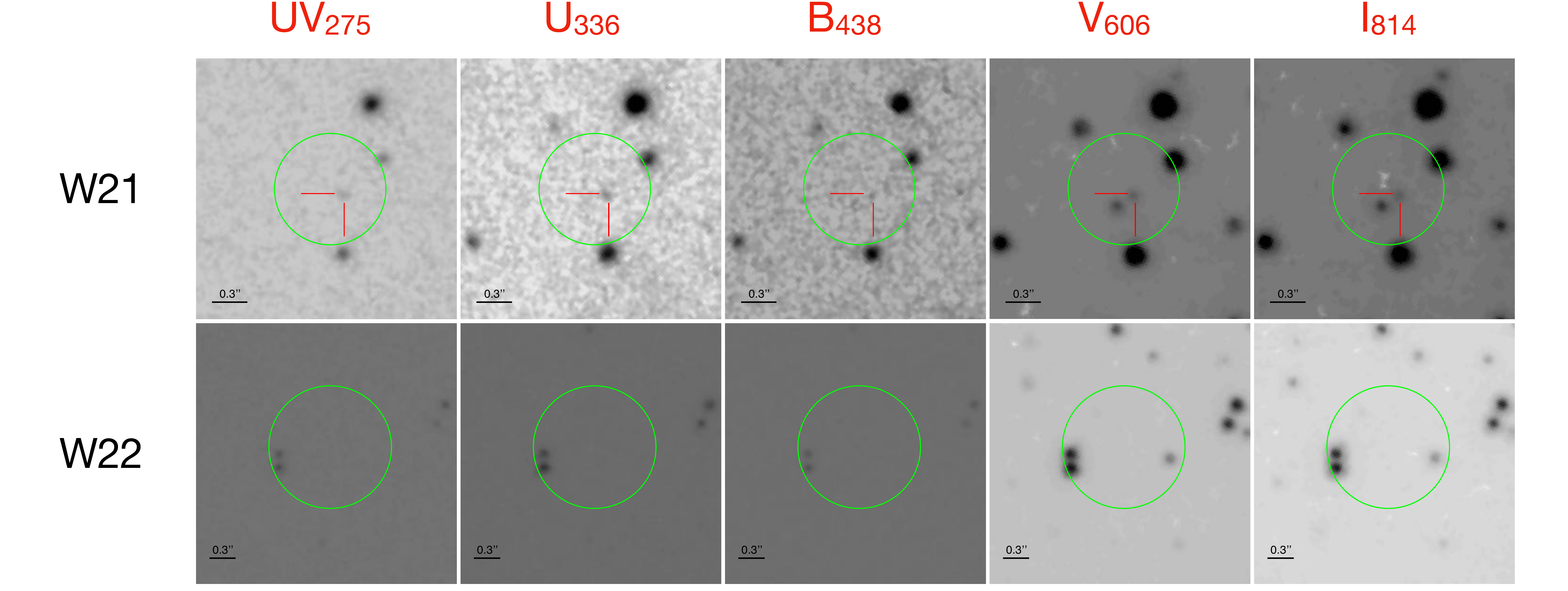}
    \contcaption{}
    \label{fig:finder_p4}
\end{figure*}


\bsp	
\label{lastpage}
\end{document}